\documentclass[10pt,onecolumn,draftclsnofoot]{IEEEtran}

\usepackage[final]{graphicx}
\usepackage[reqno]{amsmath}
\usepackage{wrapfig}
\usepackage{amssymb}
\usepackage{amsmath}
\usepackage{subfig}
\usepackage{epstopdf}
\usepackage{xcolor}
\usepackage{float}
\usepackage[paper=letterpaper,tmargin=1in,bmargin=.8in,left=.8in, right=.8in]{geometry}
\usepackage{comment}
\usepackage{dsfont}
\usepackage{cite}
\usepackage[ruled,vlined,noend]{algorithm2e}
\usepackage{algpseudocode}
\usepackage{adjustbox}
\include{pythonlisting}
\begin{document}

\sloppy
 \pagestyle{empty}
\title{Ensemble Wrapper Subsampling\\ for Deep Modulation Classification}
\author{{\large{Sharan Ramjee, {\em Student Member, IEEE}, Shengtai Ju, {\em Student Member, IEEE}, \\Diyu Yang, {\em Student Member, IEEE}, Xiaoyu Liu, {\em Student Member, IEEE},\\Aly El Gamal, {\em Senior Member, IEEE} and Yonina C. Eldar, {\em Fellow, IEEE}}}
\thanks{This work is supported in part by DARPA and AFRL under grant no. 108818, and was presented in part at the Asilomar Conference on Signals, Systems and Computers~\cite{asilomar17}.

S. Ramjee, S. Ju, D. Yang, X. Liu and A. El Gamal are with the Department of Electrical and Computer Engineering, Purdue University, West Lafayette, IN, 47907 USA (e-mail: sramjee, ju10, yang1467, liu1962, elgamala@purdue.edu).

Y. C. Eldar is with the Department of Math and Computer Science, Weizmann Institute of Science, Rechovot, Israel (e-mail: yonina@weizmann.ac.il).}
}
\maketitle

\begin{abstract}
Subsampling of received wireless signals is important for relaxing hardware requirements as well as the computational cost of signal processing algorithms that rely on the output samples. We propose a subsampling technique to facilitate the use of deep learning for automatic modulation classification in wireless communication systems. Unlike traditional approaches that rely on pre-designed strategies that are solely based on expert knowledge, the proposed data-driven subsampling strategy employs deep neural network architectures to simulate the effect of removing candidate combinations of samples from each training input vector, in a manner inspired by how wrapper feature selection models work. The subsampled data is then processed by another deep learning classifier that recognizes each of the considered 10 modulation types. We show that the proposed subsampling strategy not only introduces drastic reduction in the classifier training time, but can also improve the classification accuracy to higher levels than those reached before for the considered dataset. An important feature herein is exploiting the transferability property of deep neural networks to avoid retraining the wrapper models and obtain superior performance through an ensemble of wrappers over that possible through solely relying on any of them.
\end{abstract}

\section{Introduction}\label{sec:intro}
Automatic modulation classification plays an important role in modern wireless communications. It finds applications in various commercial and military areas. For example, Software Defined Radios (SDR) use blind recognition of the modulation type to quickly adapt to various communication systems, without requiring control overhead. In military settings, friendly signals should be securely received, while hostile signals need to be efficiently recognized typically without prior information. Under such conditions, advanced real time signal processing and blind modulation recognition techniques are required. Modulation recognition is also important for identifying the source(s) of received wireless signals, which can enable various intelligent decisions for a context-aware autonomous wireless communication system.

A typical modulation classifier consists of two steps: signal preprocessing and classification algorithms. Preprocessing tasks may include noise reduction and estimation of signal parameters such as carrier frequency and signal power. In the second step, three popular categories of modulation recognition algorithms are conventionally selected: Likelihood-Based (LB)\cite{sills1999maximum,polydoros1990detection,sapiano1996maximum,beidas1998asynchronous,panagiotou2000likelihood,hong2002antenna}, Feature-Based (FB)\cite{hsue1989automatic,hong1999identification,swami2000hierarchical,hatzichristos2001hierarchical,soliman1992signal,lichun2002comments} or using an Artificial Neural Network (ANN)\cite{mingquan1998ar,mobasseri2000digital,mingquan1996cyclic,azzouz1996modulation,nolan2001modulation}. The first compares the likelihood ratio of each possible hypothesis against a threshold, which is derived from the probability density function of the observed wave. Multiple likelihood ratio test (LRT) algorithms have been proposed: Average LRT\cite{kim1988digital}, Generalized LRT\cite{lay1994per}, Hybrid LRT\cite{hong2002antenna} and quasi-hybrid LRT\cite{sills1999maximum}. For the FB approach, the classification decision is based solely on a subset of selected features. Both LB and FB methods require precise estimates in the first step and have only been derived to distinguish between few modulation types\cite{kim1988digital,sapiano1996maximum,park2008automatic,de2010prototype}. ANN structures such as Multi-Layer Perceptrons (MLP) have been widely used as modulation type classifiers\cite{mingquan1998ar}. Traditional MLP performs well on modulation types such as AM, FM, ASK, and FSK. Recent work has shown that deep neural networks with cutting-edge structures could greatly improve the classification process (see e.g., \cite{conv} and \cite{west2017deep}), and deliver superior performance to state-of-the-art methods by enabling modulation recognition in presence of a wide variety of modulation types, and with little or no requirements from the preprocessing step.

Deep neural networks have played a significant role in the research domain of video, speech and image processing over the past few years. 
The recent success of deep learning algorithms is associated with applications that suffer from inaccuracies in existing mathematical models and enjoy the availability of large data sets. Recently, the idea of deep learning has been introduced for modulation classification using a Convolutional Neural Network (CNN) for distinguishing between 10 different modulation types~\cite{conv}. Simulation results show that a CNN not only demonstrates better accuracy results than current expert-based approaches, but also provides more flexibility in detecting various modulation types. 
Other deep neural network architectures like the Residual Network (ResNet) \cite{resnet} were also recently introduced to strengthen feature propagation in deep neural networks by creating shortcut paths between different layers in the network. By adding the bypass connections, an identity mapping is created, allowing the deep network to learn simple functions. A ResNet architecture was shown to be successful for distinguishing between 24 different modulation types in~\cite{new-resnet}. A Convolutional Long Short-term Deep Neural Network (CLDNN) was recently introduced in~\cite{CLDNN}, by combining the architectures of the CNN and the Long Short-Term Memory (LSTM) into a deep neural network and taking advantage of the complementarity of CNNs, LSTMs, and conventional deep neural network architectures. The LSTM unit is a memory unit of a Recurrent Neural Network (RNN). RNNs are neural networks with memory that are suitable for learning sequence tasks such as speech recognition and handwritten recognition. LSTM optimizes the gradient vanishing problem in RNNs by using a forget gate in its memory cell, which enables the learning of long-term dependencies. The authors in \cite{west2017deep} demonstrated the potential of LSTM units for accurately recognizing a wide range of modulation types. 

In this work, we first present three different architectures that deliver higher classification accuracy than the CNN introduced in~\cite{conv} as well as the CLDNN of \cite{west2017deep}. We design our own CNN and CLDNN architectures for the modulation recognition task, as well as derive an optimized version of the ResNet architecture of~\cite{new-resnet} by tuning the number of residual stacks. In contrast to the $75\%$ high SNR classification accuracy acheived by the CNN of \cite{conv} using the RadioML2016.10b dataset that was first considered in the same work, our CNN, CLDNN, and ResNet architectures deliver high SNR accuracy values of $83.8\%$, $88.5\%$, and $92\%$, respectively. However, we find that the performance of all these architectures, as well as the ones in~\cite{conv} and \cite{west2017deep}, suffers degradation, even at high SNR, due to confusions between similar modulation types, in particular those of QAM16 and QAM64 and those of AM-DSB and WBFM. Another major challenge facing machine learning algorithms based on deep neural network architectures is the long training time. For example, for the problem at hand, even the simple CNN architecture in \cite{conv} takes approximately 40 minutes to train using three Nvidia Tesla P100 GPU chips. This creates a serious obstacle towards the feasibility of applying such algorithms in real time, where online training is needed to adapt the network architecture to changing environmental conditions. In particular, applying deep learning to autonomous wireless communication systems anticipated in next-generation networks requires significant reduction in training time compared to state-of-the-art methods. In such systems, it is likely that training will be frequently needed to accommodate new environmental conditions. Hence, reducing training time becomes essential for the success of these algorithms. The third major challenge is hardware requirements due to sampling the received signal at high rates, which can be cumbersome in real time, particularly in wideband settings.

We tackle the three aforementioned challenges by introducing a data-driven subsampling stategy that relies on an ensemble of the three deep neural network classifiers presented in this work, as well as the ResNet as a final deep neural network classifier that recognizes the modulation type. Our strategy relies on the learning transferability property of deep neural networks, as we determine the optimal set of samples based on simulations that employ a diverse set of architectures, all of which are suitable for the considered classification task. These simulations are inspired by how wrapper feature selection methods work through model-based evaluations of feature sets. The obtained results demonstrate that not only the proposed data-driven subsampling strategy leads to significant reductions in the required training time, but it also leads to achieving unprecedented classification accuracy values and almost fully resolves the confusions - suffered by traditional methods as well as previous deep learning-based methods - between similar pairs of modulation types like QAM16 and QAM64 as well as AM-DSB and WBFM at higher SNR values (above 2 dB). Using the RadioML2016.10b dataset of \cite{conv}, the ResNet high SNR classification accuracy increases with subsampling rates as low as $\frac{1}{16}$, and goes above $99\%$ when subsampling with a rate of $\frac{1}{4}$ or higher. As further illustrated in Section \ref{sec:discussion}, we believe that subsampling led to an increase in classification accuracy in our experiments, due to its effect in reducing overfitting by projecting samples onto a lower dimensional subspace that admits a distinction between different classes through simple decision boundaries.   

The rest of this paper is organized as follows. In Section~\ref{sec:problem}, we describe the problem. We then provide a detailed description of the proposed approach in Section \ref{sec:approach}, and highlight the obtained results in Section~\ref{sec:results}. Provided by empirical evidence, we provide a detailed justification for every step of the proposed approach through a benchmarking and ablation study in Section \ref{sec:benchmark}. Finally, we provide a discussion in Section~\ref{sec:discussion} and concluding remarks in Section~\ref{sec:conclusion}.

%The rest of this document is organized as follows. In Section~\ref{sec:problem}, we describe the problem. We then detail the experimental setup in Section \ref{sec:setup}. We then analyze the performance of a Gaussian naive Bayes classifier in Section~\ref{sec:naivebayes}. This sheds light on the difficulty of the modulation classification task, and the inadequacy of traditional likelihood based techniques. In Section~\ref{sec:architectures}, we introduce deep architectures with superior performance over state of the art. We then review current subsampling and feature selection techniques in Section~\ref{sec:sota}. The data driven subsampling method along with an analysis of its performance is presented in Section~\ref{sec:dds}. Finally, we provide a discussion in Section~\ref{sec:discussion} and concluding remarks in Section~\ref{sec:conclusion}.

\section{Problem Description}\label{sec:problem}
In this work, we consider classification of the modulation type of received wireless signals, using deep neural network classifiers and subsampling techniques. We consider ten widely used modulation schemes: eight digital and two analog modulations. These consist of BPSK, QPSK, 8PSK, QAM16, QAM64, BFSK, CPFSK, and PAM4 for digital modulations, and WBFM, and AM-DSB for analog modulations. 

A general expression for the received baseband complex envelope is 
\begin{equation}\label{eq1}
r\left(t\right)=s(t;{\boldsymbol u}_\mathbf i)+n\left(t\right),
\end{equation}
where for $0\leq t\leq KT$,
\begin{equation}\label{eq2}
s(t;{\boldsymbol u}_\mathbf i)\;=\;a_ie^{j2\pi\triangle ft}e^{j\theta}{\textstyle\sum_{k=1}^K}e^{j\phi_k}s_k^{(i)}g(t-(k-1)T-\varepsilon T),
\end{equation}
is the baseband complex envelope of the received signal, and $n(t)$ is the instantaneous channel noise at time $t$. In \eqref{eq2}, $a_i$ is the received signal amplitude, $\Delta f$ is the carrier frequency offset, $\theta$ is the time-invariant carrier phase introduced by the propagation delay, $\phi_k$ is the phase jitter, $\{s_k^{(i)}, 1\leq k \leq K\}$ denotes $K$ complex symbols taken from the $i^{th}$ modulation format, $T$ represents the symbol period, $\varepsilon$ is the normalized epoch for time offset between the transmitter and signal receiver, $g(t)\;=\;P_{pulse}(t)\otimes h(t)$ is the composite effect of the residual channel with $h(t)$ denoting the channel impulse response and $\otimes$ denoting mathematical convolution, and $P_{pulse}(t)$ is the transmit pulse shape. We denote $\{a_i,\;\Delta f,\;\theta,\;\varepsilon,\;g(t),\;\{\phi_k\}_{k=1}^{K},\;\{s_k^{(i)}\}_{k=1}^K\}$ by ${\boldsymbol u}_\mathbf i$; the multidimensional vector that includes the deterministic unknown signal or channel parameters for the $i^\textrm{th}$ modulation type. \textbf{Our goal is to recognize the modulation type $i$ from a sampled version of the received signal $r(t)$}. This is achieved through a supervised machine learning algorithm that has access to labeled sample vectors. We assume that the data available for training and testing are equi-sized, and so are these available for each of the ten modulation types. We further study this problem under constraints on the allowed sampling rate. Such constraints could reflect a training time limitation, which is analyzed in this work, as well as hardware requirements (e.g., of RF sensors). 

Using the RadioML2016.10b dataset that consists of samples taken at around 6 times the Nyquist rate and 8 samples per symbol, a CNN architecture was shown to achieve $75\%$ classification accuracy at $18$ dB SNR \cite{conv}. As detailed below, we first present three deep neural network architectures that deliver state-of-the-art performance, with classification accuracy values reaching $92\%$ at high SNR. Then, we present a data-driven subsampling strategy that employs the ensemble of the presented architectures and relies on wrapper-based recursive simulations, to deliver accuracy values that exceed $99\%$ at high SNR with sampling rates around the Nyquist rate, and remain above the no subsampling accuracy with sampling rates at or above $37.5\%$ of the Nyquist rate. \textbf{To the best of our knowledge, the accuracy values obtained by applying our method with subsampling rates at or above $\frac{1}{16}$ are higher at high SNR than those obtained by applying existing methods in the literature to the same dataset, even with no subsampling, and this superior performance is uniform across the studied SNR range from -20 dB to 18 dB when applying our method with subsampling rates at or above $\frac{1}{4}$}.

\section{Designing the Ensemble Wrapper Subsampler}\label{sec:approach}
The proposed strategy utilizes training data, originally sampled at a high rate, to search for the optimal set of sample indices using an ensemble of deep neural network architectures that were found empirically to be well fit for the considered task. Once the sample indices are determined, we only sample at the corresponding times for training and testing the modulation type classifier. We will show in the sequel that samples chosen by this strategy lead to classification accuracy values that are drastically higher than the state-of-the-art. We begin by introducing three deep classifiers, each achieving high classification results on fully sampled data. Then, we build wrapper models - that we call Subsampler Nets - using each of the three architectures. We then use the ensemble of these models to build a Holistic Subsampler that exploits the diversity in performance delivered by the three models. Finally, we introduce a deterministic variant of $\epsilon$-Greedy search that finesses the obtained classification performance, by exploiting the available wrapper-based sample ranking. We present results obtained through the proposed appraoch and justify the need for each of its components in Sections \ref{sec:results} and \ref{sec:benchmark}, respectively.
\subsection{Deep Neural Network Architectures}
Our strategy employs a Convolutional Neural Network (CNN), a Convolutional Long Short-term Deep Neural Network (CLDNN), and a Residual Network (ResNet), whose details we provide below. For all architectures, we use the Adam optimizer and the categorical cross entropy loss function. We also use ReLu activation functions for all layers, except the last dense layer, where we use Softmax activation functions. \textbf{Robustness} and \textbf{diversity} were the \textbf{key design factors} that guided our choice of architectures. The former indicates that each architecture is well fit for the task, even at low sampling rates, which we verified through experimental results, and the latter indicates that the three architectures are independently trained and rely on different mechanisms for capturing task-relevant features. While a CNN relies on a fixed hierarchical representation that first extracts lower-level features through a large number of convolutional kernels, and then captures higher-level semantics through less kernels whose outputs have a wide input receptive field, the ResNet relies on shortcut connections between convolutional layers that are far apart, which provides stable training for deeper layers and allows for dynamically choosing the \emph{effective architecture} while training (see \cite[Chapters $8$-$9$]{dl-book} for more illustration). Also, the ResNet is significantly deeper than the CNN, which makes it likely to reach very different solutions. Further, unlike these two architectures, the CLDNN includes a gated LSTM layer that captures long-term temporal correlations in the convolutional output feature maps.    
\subsubsection{CNN}
\begin{figure}[H]
    \captionsetup[subfigure]{labelformat=empty}
    \centering
    \subfloat[(a)]{{\includegraphics[width=0.4\columnwidth]{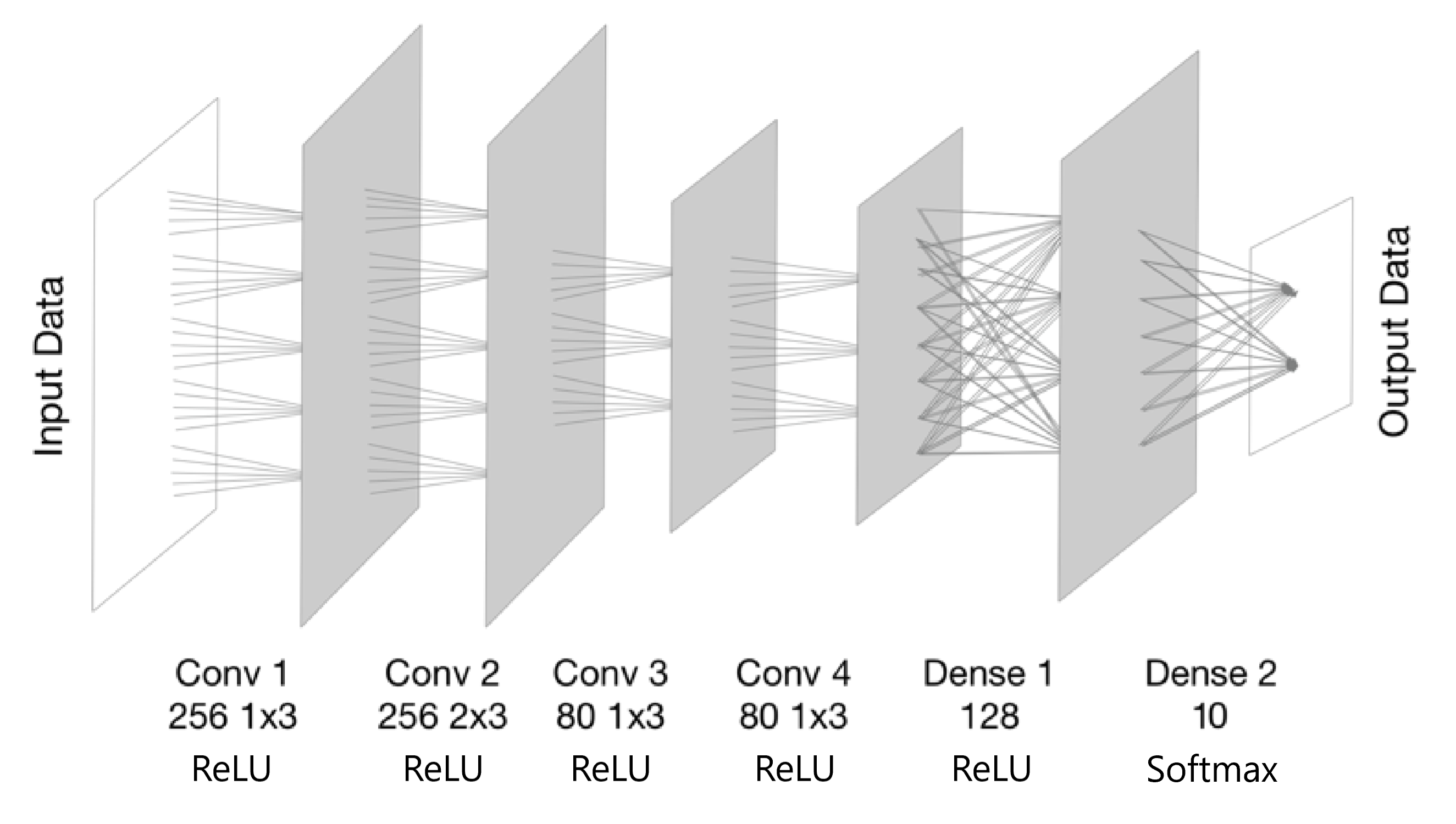}}}%
    \quad
	\subfloat[(b)]{{\includegraphics[width=0.4\columnwidth]{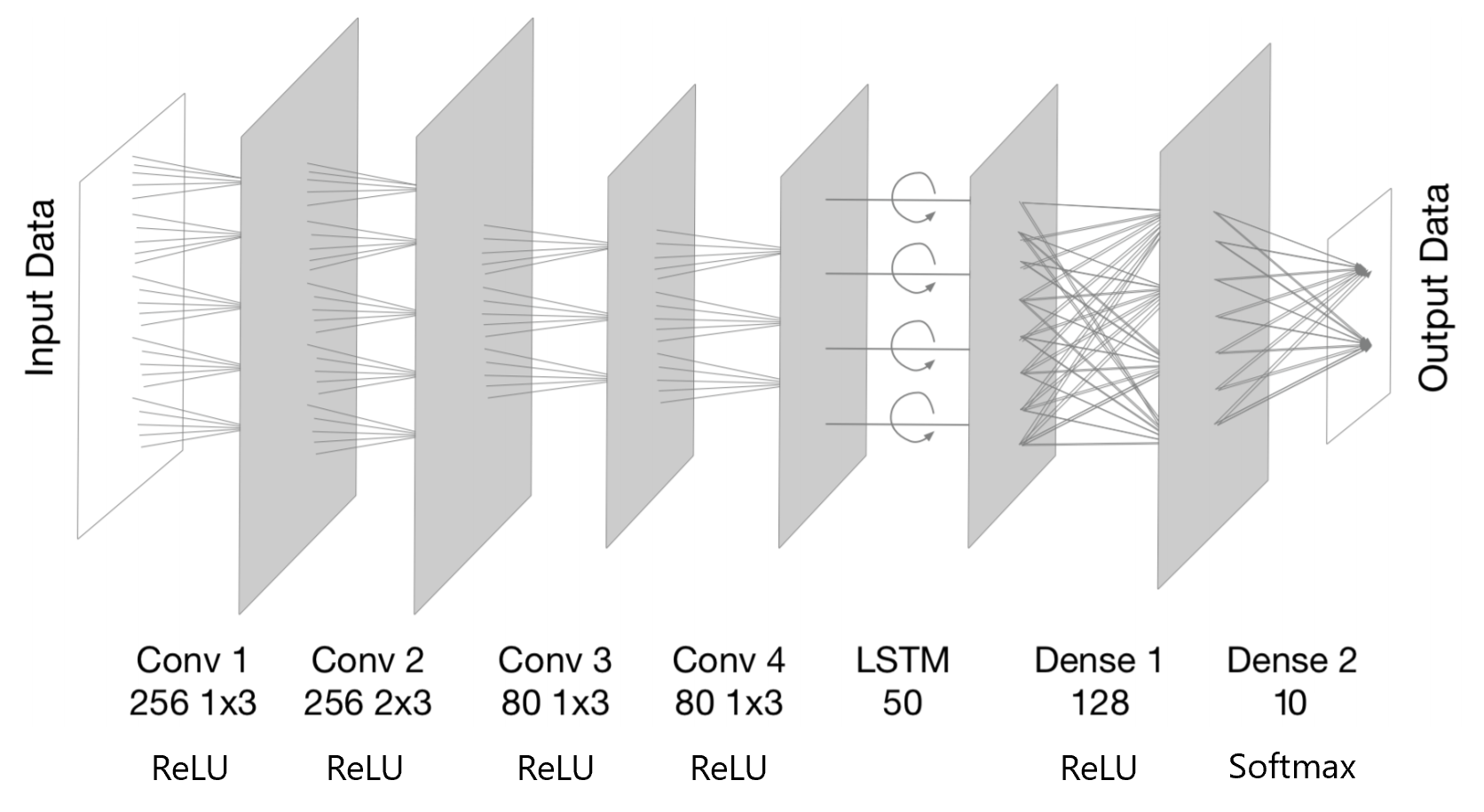}}}%
    \quad
    % \subfloat[(d)]{{\includegraphics[width=0.4\columnwidth]{resnet_arch.PNG}}}
    \caption{Architecture diagrams of (a) CNN, and (b) CLDNN.}
	\label{fig:archs}
\end{figure}
We modify the CNN2 architecture, that was proposed in \cite{conv} by having four convolutional layers, and two dense layers, as depicted in Figure \ref{fig:archs}a.
The first parameter below each convolutional layer in the figure represents the number of filters in that layer, while the second and third numbers show the size of each filter. For the two dense layers, we use 128 and 10 neurons in order of their depth in the network.

\subsubsection{CLDNN}
Inspired by \cite{CLDNN}, we proposed a CLDNN in \cite{asilomar17} by adding an LSTM layer into the CNN architecture. The detailed architecture considered for the CLDNN is shown in Figure \ref{fig:archs}b. The extra LSTM layer is placed between the convolutional layers and the dense layers. In our experiments, an LSTM layer with 50 cells provided the best accuracy.

\subsubsection{ResNet}
\begin{figure}[H]
    \centering
    \begin{tabular}{cc}
    \adjustbox{valign=b}{\subfloat[\label{tab:resnet}]{%
          \includegraphics[width=.34\columnwidth]{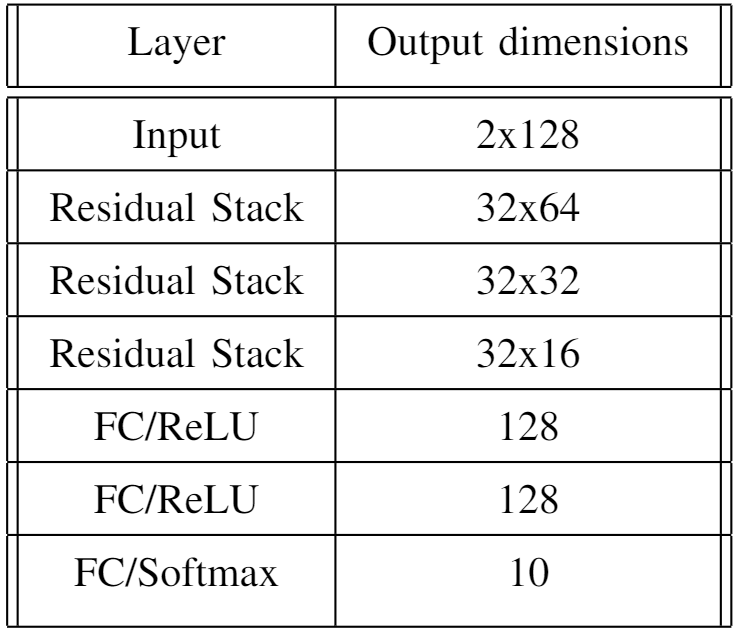}}}
    &      
    \adjustbox{valign=b}{\begin{tabular}{@{}c@{}}
    \subfloat[\label{subfig-2:dummy}]{%
          \includegraphics[width=.25\columnwidth]{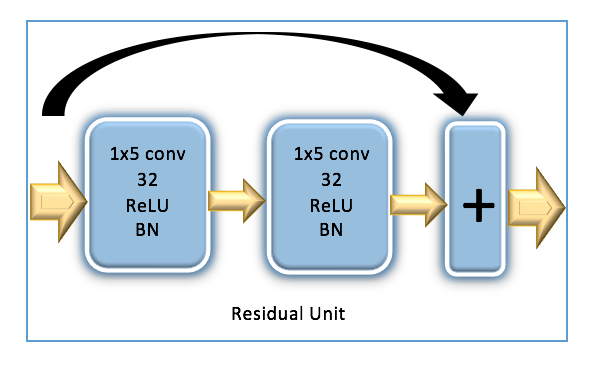}} \\
    \subfloat[\label{subfig-3:dummy}]{%
          \includegraphics[width=.25\columnwidth]{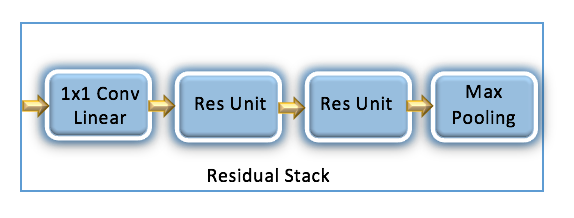}}
    \end{tabular}}
    \end{tabular}
    \caption{(a): The Resnet architecture, (b) A Residual unit, and (c) A Residual stack.}
    \label{fig:resnet_arch}
  \end{figure}

As neural networks grow deeper, their learning performance is challenged by problems like vanishing or exploding gradient and overfitting, and therefore both training and testing accuracy start to degrade after the network reaches a certain depth. The Deep Residual Network (ResNet) architecture that was introduced in the ImageNet and COCO 2015 competitions\cite{resnet}, resolves accuracy degradation issues in deeper neural networks and has been shown to be a robust choice for a wide range of machine learning tasks. 
Inspired by the ResNet architecture in \cite{new-resnet}, we designed a similar ResNet but with three residual stacks instead of six, as we found that choice to lead to increased classification accuracy. In our network, three residual stacks are followed by three fully connected layers, where each residual stack consists of one convolutional layer, two residual units, and one max-pooling layer. As seen in \cite{new-resnet}, for each residual unit, a shortcut connection is created by adding the input of the residual unit with the output of the second convolutional layer of the residual unit. Finally, each convolutional layer in the residual unit uses a filter size of 1x5 and is followed by a batch normalization layer for optimization stability. The overall architecture is observed in Figure \ref{fig:resnet_arch}. As we illustrate below, this architecture delivered the best - or very close to the best - performance among all considered architectures for a wide range of SNR values that only excludes extremely low values.

\subsection{Subsampler Nets: A Wrapper Feature Selection Approach}\label{sec:subsampler_nets}
\begin{figure}[H]
    \centering
	\includegraphics[width=0.7\columnwidth]{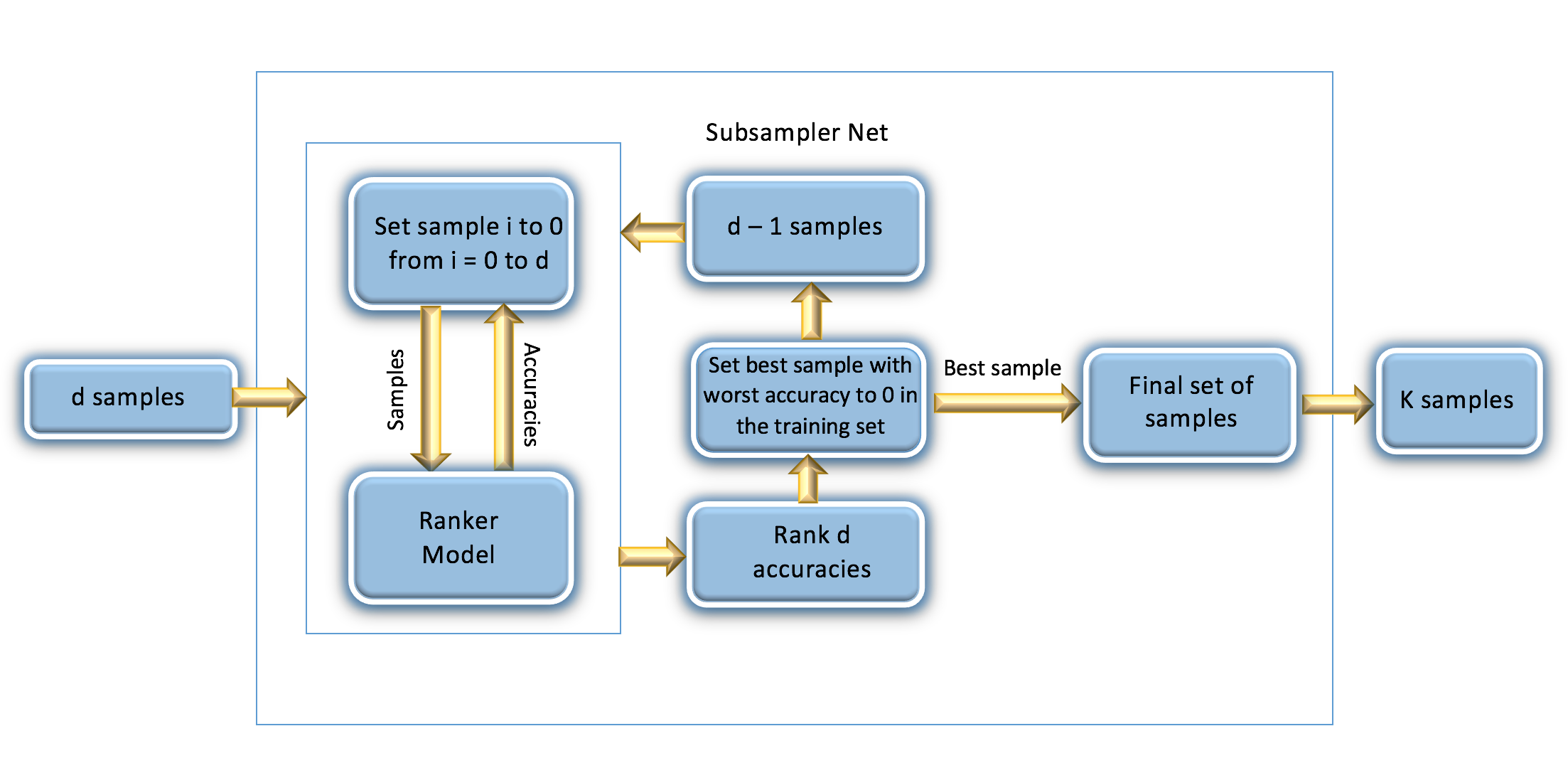}
	\caption{Subsampler Net that chooses $k$ representative samples out of a pool of $d$ samples.}
	\label{fig:subsampler_net}
\end{figure}
The first building block in our ensemble method that employs the architectures provided above, is a supervised wrapper feature selection algorithm that we call the \textbf{Subsampler Net}, which uses a deep neural network to rank the importance of each sample in accordance to the relative drop in classification accuracy that occurs when that sample is removed (set to 0). Similar to other wrapper feature selection methods, a Subsampler Net relies on a classifier to rank sample importance. We will refer to this classifier as the \textbf{Ranker Model}. We use pre-trained models for each of the above three architectures as Ranker Models to construct three separate Subsampler Nets.

Suppose we want to obtain $k$ samples from a pool of $d$ samples. As shown in Figure \ref{fig:subsampler_net}, the Subsampler Net first starts by setting a sample to $0$ (setting both the real and imaginary parts of that sample to $0$) in a batch of training validation examples and evaluating it using the Ranker Model, which will then provide us with a classification accuracy for that batch. Setting a sample to $0$ means that the input neurons to the model for the two features corresponding to that complex sample are dead, which allows us to simulate the removal of that sample from the signal as all the weights from these neurons in the input layer will not contribute to the outcome of the model. This is done $d$ times by setting each of the $d$ samples to $0$. After evaluating each of the $d$ samples, we are left with $d$ classification accuracies that correspond to the ability of the model to classify the signal if each of the samples were to be removed. The most important sample, which is the sample whose removal results in the lowest classification accuracy, is then permanently set to $0$ for this batch of training examples and added to the final set of samples. Now, we are left with $d-1$ samples and this process is repeated until we are finally left with $k$ samples. The Subsampler Net construction is detailed in Algorithm \ref{algo:subsampler_net}.

\begin{algorithm}[H]
\SetAlgoNoLine
\DontPrintSemicolon
%\begin{algorithmic}[1]
\vspace{0.05 in}
\textbf{Inputs}: $k$: Final No. of samples; trainSet: Training Dataset; rankerModel: Trained Model that ranks samples

\textbf{Outputs}: sampleList: List of $k$ selected sample indices
%\\[0.1in]
\begin{algorithmic}
\Function{subsamplerNet}{$k$, trainSet, rankerModel}
    \State Initialize sampleList to an empty list
    \State \For{$i = 0$ to $k$}{%\\[0.05in]
        \State Initialize accList to an empty list
        \State Set candidateList as set of sample indices not in sampleList
        \State \For{$j$ in candidateList}{\vspace{0.05 in}
            \State Set sample with index j to 0 in trainSet
            \State accuracy = rankerModel(trainSet)
            \State Append ($j$, accuracy) to accList
            \State Set sample with index j back to original value\vspace{0.05 in}
        }\State \textbf{end}
        \State Sort accList by order of increasing accuracy
        \State Append sample index with lowest accuracy in accList to sampleList
        \State Permanently set this sample to $0$ for all examples in trainSet%\vspace{0.1 in}
    \State \textbf{return} (sampleList)}\vspace{0.05 in}
\EndFunction\vspace{0.05 in}
\end{algorithmic}
\caption{Subsampler Net}
\label{algo:subsampler_net}
\end{algorithm}
While attempting this method, we found that normalizing the data by setting the mean of each sample to $0$ and the variance to $1$ improves performance because when we set an input sample to $0$, we are effectively setting it to the mean, and lower variances now manifest as lower weights in the input layer \cite{Ramjee2019EfficientWF}. We also observed, from experiments that rely on a discrete set of SNR values, that the sample indices chosen for batches of validation examples belonging to the same SNR value were the same in most cases, while they were likely to be distinct from those chosen for batches belonging to different SNR values. Therefore, we divide the available data according to SNR value and obtain a set of $k$ sample indices for each SNR.

\subsection{The Holistic Subsampler: The Best of All Worlds}\label{sec:holistic}
We next introduce the notion of the Holistic Subsampler, which combines the ability of all three Ranker Models in order to capitalize on their diversity of performance. After the set of samples, that match the required sampling rate, are collected for each of the three Ranker Models, we divide these samples into three tiers.
\begin{wrapfigure}{r}{5 cm}
%    \centering
    \includegraphics[width=5 cm]{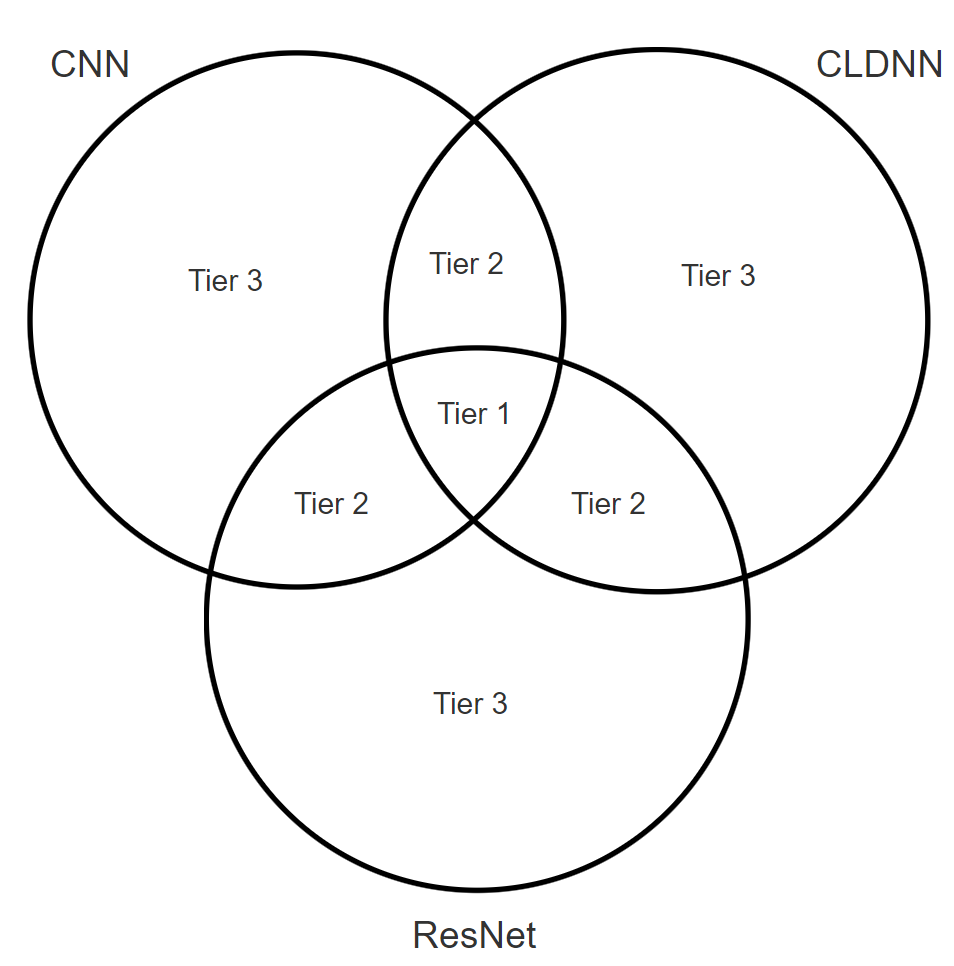}
    \caption{Tier Division.} 
    \label{fig:venn_diagram}
\end{wrapfigure}
The Venn diagram in Figure \ref{fig:venn_diagram} illustrates the division of the tiers. Tier 1 consists of the intersection of all three sets of samples, Tier 2 consists of the samples that belong to two of the three sets of samples, and Tier 3 consists of samples that are selected by only one Subsampler Net. The samples within each of the tiers are sorted according to the sum of the classification accuracy values that occur when that sample is removed using the corresponding Ranker Models. For example, for Tier 2 samples, we sort the samples using the sum of the two obtained classification accuracy values when the sample is removed (set to 0). Akin to the Subsampler Net, the Holistic Subsampler is a recursive algorithm that first selects the best sample, then sets the value of the corresponding sample index to 0 for the whole training set, and calls itself again to find the next best sample. This is done until the desired number of samples $k$ is reached. To find the best sample, the top sample - corresponding to the lowest sum of classification accuracy values - is selected from Tier 1. If Tier 1 is empty, then the top sample from Tier 2 is selected. If Tier 1 and Tier 2 are both empty, then the top sample from Tier 3 is selected.

\subsection{$\epsilon$-Greedy Search: The Final Piece of The Puzzle}\label{sec:egreedy}
We note that the Subsampler Net merely selects the best sample at each iteration, without regard to how the selection of a subsequent sample will affect the importance of the currently selected sample to the classification task. We chose to do this in the interest of saving training time of Subsampler Nets in order to render the implementation of the proposed method feasible using low-power communication devices. To tackle this problem, we next propose a variant of the $\epsilon$-greedy algorithm \cite{stadie2015incentivizing} in order to explore candidate combinations for subsequent best samples while taking into account dependence relationships between the selected samples.

\begin{algorithm}[H]
\SetAlgoNoLine
\DontPrintSemicolon
\vspace{0.05 in}
\textbf{Inputs}: d: Total No. of samples; k: Final No. of samples; $\epsilon$: exploration factor that is the fraction of total samples to be explored; prevSnrAcc: Classification accuracy at the preceding SNR value; trainSet: Training Dataset\\
\textbf{Outputs}: finalIdx: Set of sample indices whose removal leads to the lowest accuracy among combinations explored; finalSNRAcc: Accuracy obtained using trainSet when the sample indices in finalIdx are selected
\vspace{0.1 in}
\begin{algorithmic}
\Function{$\epsilon$-Greedy}{$k$, $\epsilon$, prevSnrAcc, trainSet}
    \State Call SubsamplerNet using CNN, CLDNN, and ResNet as the Ranker Models
    \State Set sampleIdx as the ordered set of the $k$ sample indices selected by the Holistic Subsampler
    %\State Set sampleAcc as the set of lowest $k$ accuracies using the Holistic Subsampler
     \State Set currSnrAcc as accuracy of ResNet architecture when trained with selected samples
    \State \If{$k=0$}{\vspace{0.05 in}
        \State \textbf{return} (sampleIdx, currSnrAcc) if currSnrAcc $>$ prevSnrAcc
        \State \textbf{return} (NULL, NULL) otherwise
    }
    \State \Else{
        \State \For{$i = 0$ to $\min(k,\epsilon d)$}{\vspace{0.05 in}
            \State Set trainSet[sampleIdx[$i$]] to 0
            \State Set (finalIdx, finalSnrAcc) = \Call{$\epsilon$-Greedy}{$k-1$, $\epsilon$, prevSnrAcc, trainSet}
            \State Set trainSet[sampleIdx[$i$]] back to original value
            \State Add sampleIdx[$i$] to finalIdx
            \State \textbf{return} (finalIdx, finalSnrAcc) if returned values are not NULL \vspace{0.1 in}
        }
        \State \textbf{done}
        \State \textbf{return} (NULL, NULL)
        \vspace{0.1 in}
    }
    \State \textbf{end}
\EndFunction \vspace{0.05 in}
\end{algorithmic}
\caption{$\epsilon$-Greedy Search}
\label{algo:e-greedy}
\end{algorithm}

According to Algorithm \ref{algo:e-greedy}, we introduce $\epsilon$, the exploration factor that determines the number of candidate samples considered for selection at each step. If $\epsilon=0.1$, then in every step, we explore the $10\%$ best samples according to the ranking provided by the Holistic Subsampler. This is unlike the conventional $\epsilon$-greedy algorithm, where $\epsilon$ represents the probability that the decision taken deviates from the top greedy choice. Our variant of the algorithm explores all of the top routes and $\epsilon$ is the parameter that determines the number of top routes explored. The Multi-Armed Bandit problem \cite{bubeck2012regret}, which is one of the most popular applications of the traditional $\epsilon$-Greedy Algorithm, is based on a scenario where there is a single top choice and all the other choices are of the same importance before exploration. However, in our case, the choices apart from the top choice have a ranking that reflects their relative importance. Therefore, unlike the Multi-Armed Bandit problem, we do not need to waste time exploring unknown paths, and this is the rationale behind our deviation from the conventional $\epsilon$-Greedy Algorithm.

\begin{figure}
    \centering
	\includegraphics[width=0.7\columnwidth]{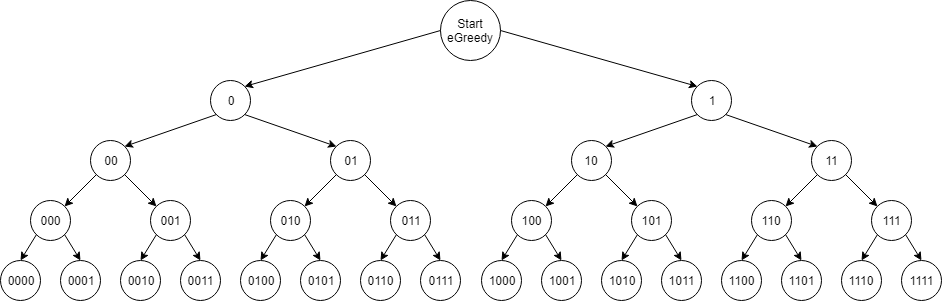}
	\caption{Illustration of $\epsilon$-Greedy Search with $\epsilon=\frac{2}{d}$, where $d$ is the total number of available samples, and a subsampling rate of $\frac{4}{d}$. A 0 corresponds to removing the best sample, while a 1 corresponds to removing the second best sample, according to the ranking of the Holistic Subsampler. The samples are re-ranked after each removal.}
	\label{fig:egreedy_decision_tree}
\end{figure}
As observed in Figure \ref{fig:egreedy_decision_tree}, the $\epsilon$-Greedy Search can be represented as a traversal algorithm over an $\epsilon d$-ary tree whose depth is equal to the desired number of samples $k$, where the subsampling rate is $\frac{k}{d}$. Each node in the tree corresponds to the selection of a combination of sample indices. The nodes at the same depth are arranged by increasing order of accuracy when removed, which implies that the combination of samples corresponding to the left child of a node has higher priority than that corresponding to the right child of the same node, for the case when $\epsilon d=2$.
The root node does not represent any sample and is added just for the sake of illustration of how the tree is formed. The leaf nodes are searched from left to right until a classification accuracy that is satisfactory is reached. 

We note that setting $\epsilon=\frac{1}{d}$ is equivalent to having an unaltered Holistic Subsampler that greedily chooses the best sample at each iteration, which leads to the leftmost leaf of the tree in Figure~\ref{fig:egreedy_decision_tree}. This corresponds to node $0000$ because only a tree with a single branch is formed with the nodes 0, 00, 000, and 0000. Increasing the value of $\epsilon$ expands the scope of exploration.

\subsection{Ensemble Wrapper Subsampling}\label{sec:everything}
We here finalize the specification of the proposed approach. Given input data with $d$ samples per example, we first initialize $\epsilon$ to $\frac{1}{d}$ and invoke the $\epsilon$-Greedy Search. As mentioned earlier, this is the same as invoking the Holistic Subsampler on its own. Next, we proceed to the next SNR value available in the training set. 
The $\epsilon$-Greedy Search is invoked with an $\epsilon$ value of $\frac{1}{d}$ for this next training set belonging to the next SNR value. If the accuracy is lower than the accuracy for the previous SNR value, then $\epsilon$-Greedy Search is invoked again after doubling the $\epsilon$ value to $\frac{2}{d}$. The $\epsilon$-Greedy Search stops exploring once the accuracy is greater than the accuracy obtained for the previous SNR value. We repeatedly double $\epsilon$ and invoke $\epsilon$-Greedy Search until this stopping criterion is met. The pseudocode for this strategy is given in Algorithm \ref{algo:main}.

The function described in Algorithm \ref{algo:main} returns the selected set of $k$ sample indices for each SNR value. Note that douling the value of $\epsilon$ for the $\epsilon$-Greedy Search corresponds to searching for combinations of sample indices in a tree that has twice the arity.
\begin{algorithm}
\SetAlgoNoLine
\DontPrintSemicolon
\vspace{0.05in}
\textbf{Inputs}: d: Total No. of samples; k: Final No. of samples; trainSet: Training Dataset\\
\textbf{Outputs}: idxDict: Dictionary with SNR as key and sets of $k$ sample indices each as values
\vspace{0.1in}
\begin{algorithmic}
\Function{EnsembleWrapperSubsampler}{$k$, trainSet}
    \State Divide trainSet based on SNR
    \State Initialize idxDict as an empty dictionary
    \State Initialize prevSnrAcc to 0
    \State \For{snrValue in set of SNR values}{\vspace{0.05 in}
        \State Initialize $\epsilon = \frac{1}{d}$
        \State Initialize snrIdx as NULL
        \State \While{snrIdx is NULL}{\vspace{0.05 in}
            \State Set (snrIdx, snrAcc) = \Call{$\epsilon$-Greedy}{k, $\epsilon$, prevSnrAcc, trainSet}
            \State Set $\epsilon = 2\epsilon$\;
            \textbf{done}\vspace{0.1 in}
        }
        \State Set snrIdx as the value to snrValue key in idxDict
        \State Set prevSnrAcc as snrAcc\;
        \textbf{done}\vspace{0.1 in}
    }
    \State \textbf{return} idxDict
\EndFunction\vspace{0.05 in}
\end{algorithmic}
\caption{Ensemble Wrapper Subsampler}
\label{algo:main}
\end{algorithm}

\section{Experimental Results}\label{sec:results}
In this section, we present an experimental evaluation of the proposed method. First, we specify the dataset used, the programming environment, and hyperparameter settings, then we present the obtained classification accuracy results while highlighting important insights, and finally quantify the reduction in training time for the final classifier with different subsampling rates. 
\begin{figure}[H]
\centering
	\includegraphics[width=5in]{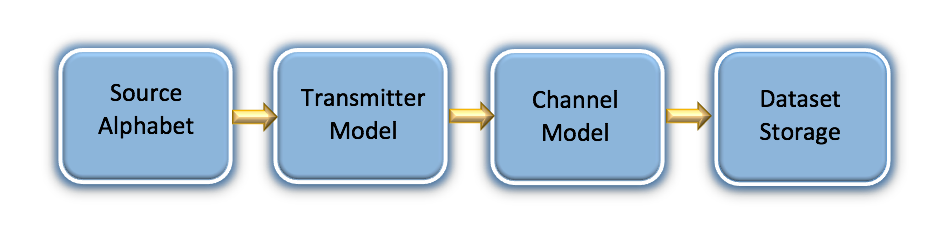}
	\caption{High-level framework for data generation.}
	\label{fig:frame}
\end{figure}
\subsection{Dataset}\label{sec:dataset}
We use the RadioML2016.10b synthetic dataset generated in~\cite{conv} as the input data. Details about the generation of this dataset can be found in \cite{datagen}. Figure \ref{fig:frame} shows a high-level framework for the data generation process. For digital modulations, the entire Gutenberg works of Shakespeare in ASCII is used, with whitening randomizers applied to ensure equiprobable symbols and bits. For analog modulations, a continuous voice signal consisting of acoustic voice speech with some interludes and off times is used as input. The modulation rate is 8 samples per symbol.
\begin{figure}[H]
    \centering
	\includegraphics[width=0.5\columnwidth]{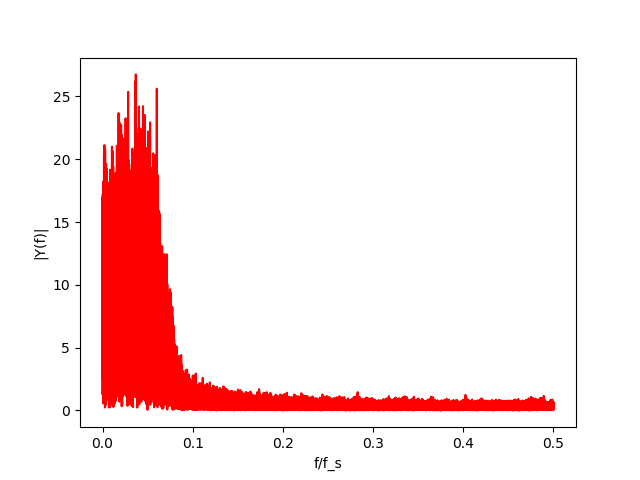}
	\caption{One-sided normalized FFT for a BPSK signal. A value of $0.5$ on the horizontal axis corresponds to the Nyquist rate (Bandwidth is half the sampling rate). Most of the signal energy is within a band of around $\frac{1}{12}$ of the sampling rate.}
	\label{fig:fft}
\end{figure}
The dataset is split equally among all ten considered modulation types.
For the channel model, physical environmental noises like thermal noise and multipath fading were simulated. The models for generating random channel and device imperfections, that determine the parameters in \eqref{eq2}, are detailed in~\cite{datagen}\footnote{Dataset generation parameters are also available at https://github.com/radioML/dataset}. When packaging data, the output stream of each simulation is randomly segmented into vectors as the original dataset with a sample rate of 1M samples per second. Similar to the way that an acoustic signal is windowed in voice recognition tasks, a sliding window extracts 128 samples with a shift of 64 samples, which forms a sample vector in the dataset. 160,000 sample vectors generated using the GNU-radio library developed in~\cite{datagen} are segmented into training and testing datasets. Each example consists of 128 samples, that are represented as a 2$\times$128 vector with real and imaginary parts separated. The SNR of the samples is uniformly distributed from -20 dB to 18 dB, with a step size of 2 dB, i.e., the dataset is equally split among all SNR dB values in $\{-20,-18,-16,\cdots,18\}$.

We note from the frequency domain representation of the input waveform depicted in Fig. \ref{fig:fft} that the sampling rate of the input waveform is around 6 times the Nyquist rate. 
\subsection{Implementation Details}
We used Keras with TensorFlow as a backend, and a GPU server equipped with three Tesla P100 GPUs with 16 GB memory. For all architectures, we used a batch size of 1024, and a learning rate of 0.001. \textbf{Only the training set is used by the subsampling algorithm} described in Section \ref{sec:approach} with a validation split of 0.25. After selecting the set of sample indices for each of the 20 considered SNR values, we train the ResNet classifier with the corresponding samples, as we found it to deliver the best performance among the three considered architectures\footnote{Code is available at: https://github.com/dl4amc/dds}.

\begin{figure}
    \captionsetup[subfigure]{labelformat=empty}
    \centering
    \subfloat[(a)]{{\includegraphics[width=0.40\columnwidth]{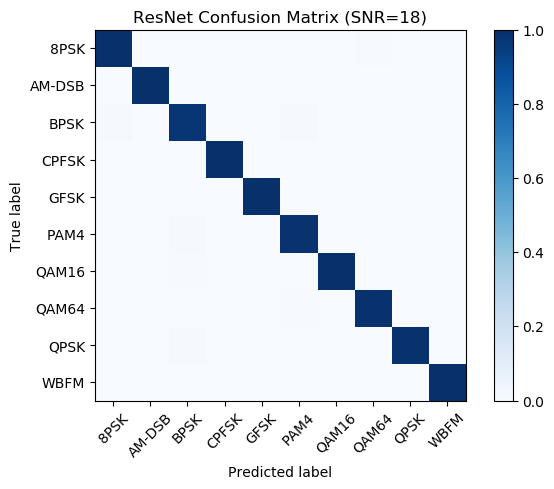}}}%
    \qquad 
    \subfloat[(b)]{{\includegraphics[width=0.55\columnwidth]{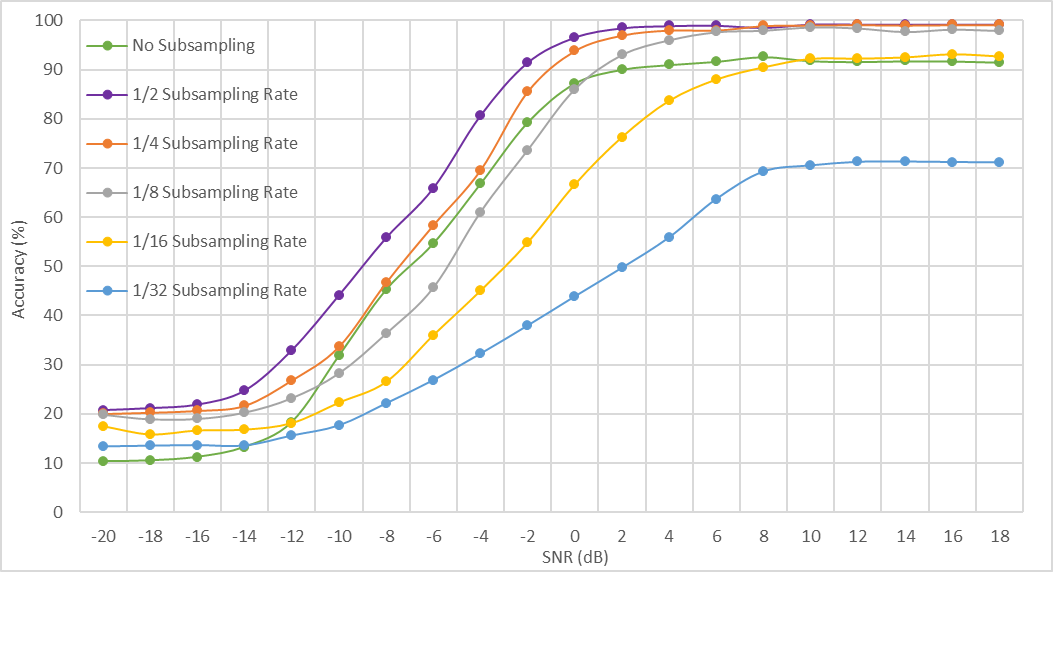}}}
    \caption{(a) ResNet Confusion Matrix after Ensemble Wrapper Subsampling using a subsampling rate of $\frac{1}{2}$ at $18$ dB SNR.\\ (b) Accuracy vs SNR for ResNet with Ensemble Wrapper Subsampling.}
	\label{fig:after_dds_plot}
\end{figure}
\subsection{Classification Accuracy}
We present the obtained results at different sampling rates in Figure \ref{fig:after_dds_plot}. From our results, we note the following.
\begin{itemize}
    \item \textbf{Subsampling can lead to higher accuracy:} Applying the proposed ensemble wrapper subsampling strategy can result in dramatic improvements in classification accuracy, particularly at low SNR values. The direct cause for this phenomenon at high SNR is resolving confusions between the QAM16/QAM64 and AM-DSB/WBFM pairs, as we illustrate in Section \ref{sec:benchmark}. To the best of our knowledge, state-of-the-art methods fail at resolving these confusions. We believe that this is because our subsampling strategy reduces overfitting, as we elaborate in Section \ref{sec:subsampling_high_acc}.
    
    \item \textbf{Sub-Nyquist Sampling:} As noted above, the considered data is originally sampled at around 6 times the Nyquist rate. A subsampling rate of $\frac{1}{16}$ hence corresponds to around $37.5\%$ of the Nyquist rate, and leads to slightly higher classification accuracy at very high SNR and significantly higher accuracy at very low SNR, than the case with no subsampling. This observation could carry important implications in practice, as the sampling hardware requirements can be dramatically simplified (see \cite{yonina-book} for more illustration).
    
    \item \textbf{Minimal Sample Set:} Based on the loss in classification accuracy, we can choose the smallest set of samples (smallest value of $k$) that gives us a classification accuracy higher than a given classification accuracy requirement. For instance, $20$ is the smallest number of samples (around Nyquist rate) that can be selected such that the classification accuracy has to be higher than $99\%$ at $18$ dB SNR. Similarly, $8$ is the smallest number of samples $\left(\text{around }37.5\% \text{ of Nyquist rate}\right)$ that can be selected given that the classification accuracy has to be higher than $90\%$ at $18$ dB SNR.
\end{itemize}

\subsection{Training Time}
As a result of subsampling, the training time of the classifier is reduced due to the reduced input dimensions. We show the reduction in training times and high SNR classification accuracy of the ResNet classifier for different subsampling rates in Table \ref{tab:training_times}. Note that \textbf{a subsampling rate as low as $\frac{1}{16}$, which corresponds to the sub-Nyquist regime with around $37.5\%$ of the Nyquist rate, and results in approximately $\frac{1}{3}$ of the original training time, still results in a classification accuracy higher than that without subsampling}.
\begin{table}[H]
\begin{center}
\caption{Comparison of Training Time and high SNR accuracy for the ResNet after Ensemble Wrapper Subsampling.}
\label{tab:training_times}
 \begin{tabular}{||c | c | c | c | c ||}
\hline
Samples     &Time per Epoch     &Epochs     &Total Training Time     &Accuracy ($18$ dB SNR) \\
\hline\hline
 All        &32.0642s   &37     &1186.3754s      &91.49\% \\
\hline
 1/2        &26.9615s   &33     &889.7295s       &99.27\% \\
\hline
 1/4        &24.2615s   &29     &703.5835s       &99.13\% \\
\hline
 1/8        &21.8361s   &27     &589.5747s       &97.94\% \\
\hline
 1/16       &17.1167s   &26     &445.0342s       &92.67\% \\
\hline
 1/32       &11.6828s   &24     &280.3827s       &71.14\% \\[1ex]
\hline
\end{tabular}
\end{center}
\end{table}
\section{Benchmarking and Ablation Study}\label{sec:benchmark}
Supported by experimental results, we first provide an analysis of our proposed approach with regard to traditional approaches, and then provide a justification for each of its components. We begin by motivating the need for deep learning via analyzing the performance of a Bayes classifier. Then, we present the results obtained with the considered deep neural network architectures with no subsampling, and highlight the modulation type pairs that are difficult to distinguish even at high SNR. Having motivated the need for subsampling, we then compare the results obtained through our proposed approach with conventional subsampling and feature selection schemes. We finally present an ablation study to demonstrate the performance degradation caused when removing any of the components of the proposed method.

\subsection{Gaussian Naive Bayes Classifier}\label{sec:naive}
We first discuss the performance of a Gaussian naive Bayes classifier, to assess the difficulty of the considered modulation classification task. It will then become clear how deep learning significantly outperforms the naive Bayes classifier. The Gaussian naive Bayes classifier can be described through the conditional probabilities:
\begin{equation}
    P(x_i|y) = \frac{1}{\sqrt{2\pi\sigma_y^2}} e^{-\frac{(x_i-\mu_y)^2}{2\sigma_y^2}},
    \label{eq:lrt}
\end{equation}
where $P(x_i|y)$ is the likelihood of an observed instance $x_i$ belonging to a certain class $y$, $\sigma_y^2$ is the observed variance of class $y$, and $\mu_y$ is the observed mean of class $y$. The predicted output of $x_i$ is the class that maximizes the likelihood function. Instead of trying to classify all ten modulation types, we only used certain pairs to further demonstrate the performance of the Gaussian naive Bayes classifier for simpler tasks. 

\begin{table}[H]
\begin{center}
\caption{Gaussian naive Bayes classifier results for different modulation pairs at all SNRs and at $18$ dB SNR.}
\label{table:naive}
 \begin{tabular}{||c | c | c||} 
 \hline
 Modulation Pairs & Classification Accuracy (All SNR) & Classification Accuracy (18 dB) \\ [0.5ex] 
 \hline\hline
 PAM4 vs QAM64 & 68.5\% & 82\% \\
 \hline
 QAM16 vs QAM64 & 50\% & 51\% \\
 \hline
 AM-DSB vs BPSK & 70\% & 81\% \\
 \hline
 AM-DSB vs WBFM & 53\% & 53\% \\
 %\hline
 %PAM4 vs QAM64 ($18$ dB) & 82\% \\
 %\hline
 %QAM16 vs QAM64 ($18$ dB) & 51\% \\
 %\hline
 %AM-DSB vs BPSK ($18$ dB) & 81\% \\
 %\hline
 %AM-DSB vs WBFM ($18$ dB) & 53\% \\[1ex]
 \hline
\end{tabular}
\end{center}
\end{table}
From the results in Table \ref{table:naive}, we note that even when the Bayes classifier is trained to distinguish pairs that are not challenging at the maximum SNR value, its maximum accuracy is $82\%$. Further, for challenging pairs, the performance is similar to random guessing even at high SNR.

\subsection{Deep Learning with no Subsampling}\label{sec:nosubsampling}
\begin{figure}[H]
    \captionsetup[subfigure]{labelformat=empty}
    \centering
    \subfloat[ResNet confusion matrix at 18 dB SNR.]{{\includegraphics[width=0.40\columnwidth]{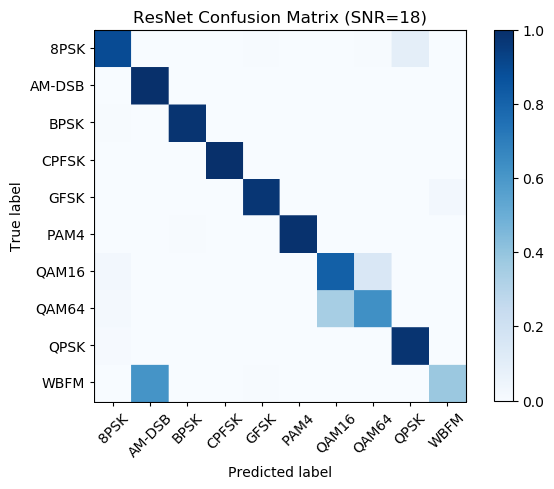}}}%
    \qquad 
    \subfloat[Accuracy vs SNR for each of the architectures.]{{\includegraphics[width=0.55\columnwidth]{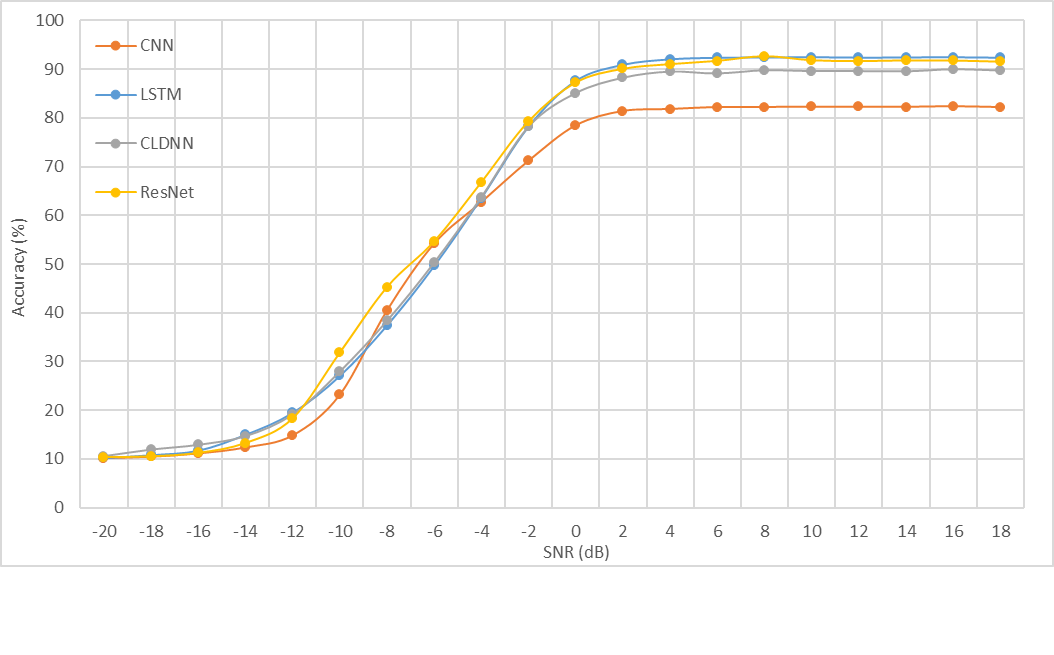}}}
    \caption{Baseline results with no subsampling for modulation classification.}
	\label{fig:before_dds_plot}
\end{figure}

We present in Figure \ref{fig:before_dds_plot} the classification accuracy of the considered architectures using the considered dataset with no subsampling. We note that - similar to previous work on deep learning for modulation classification - most of the misclassifications at high SNR are due to confusions between the AM-DSB/WBFM and the QAM16/QAM64 pairs, which is evident from the ResNet 18 dB confusion matrix depicted in Figure \ref{fig:before_dds_plot}a. We observe by comparing to Figure \ref{fig:after_dds_plot} how the proposed data-driven subsampling method leads deep neural network classifiers to clear this confusion. We believe that this is due to overfitting reduction, as further illustrated in Section \ref{sec:discussion}.  We further note that we also considered a pure LSTM architecture by fine tuning that of~\cite{lstm} for the task. Even though this architecture delivered good performance with no subsamlping as shown in Figure \ref{fig:before_dds_plot}b, we chose not to use it in our proposed method as it suffered drastic performance degradation with subsampling. We believe that this is due to extreme sensitivity of the captured temporal correlations to absence of few samples. 

\subsection{Conventional Subsampling}\label{sec:conv}
We provide in Figure \ref{fig:resnet_comparisons}a a comparison between the proposed approach and four different subsampling techniques; namely: 1) Uniform Subsampling: where a sample is taken every fixed amount of time, 2) Random Subsampling: where the indices of selected samples is determined randomly with equal probabilities, 3) Magnitude Rank Subsampling: where the indices corresponding to samples with top magnitude values in each example are selected, and 4) Principal Component Subsampling (PCS), where first Principal Component Analysis (PCA) is done over the training set, and the indices corresponding to samples with the top PCA coefficient total magnitude values are selected. A more thorough explanation of these methods is available in \cite{ramjee2019fast}. We note that the proposed method leads to - uniformly across all SNR values - superior performance than all these methods. The figure demonstrates results slightly below the Nyquist rate, but this observation extends to all considered subsampling rates. In particular, random subsampling delivers better performance that the other three methods at lower rates, which agrees with the intuition in \cite{yonina-book}, but this performance remains lower than that of the proposed approach. 

\subsection{Conventional Feature Selection}\label{sec:convfeat}

\begin{figure}[H]
    \captionsetup[subfigure]{labelformat=empty}
    \centering
    \subfloat[(a)]{{\includegraphics[width=0.45\columnwidth]{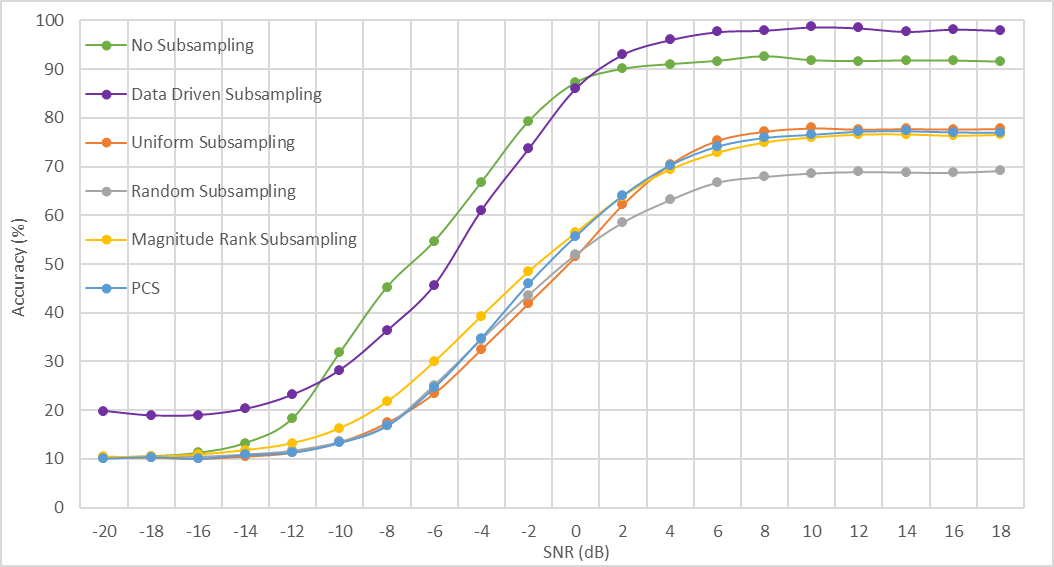}}}%
    \qquad 
    \subfloat[(b)]{{\includegraphics[width=0.45\columnwidth]{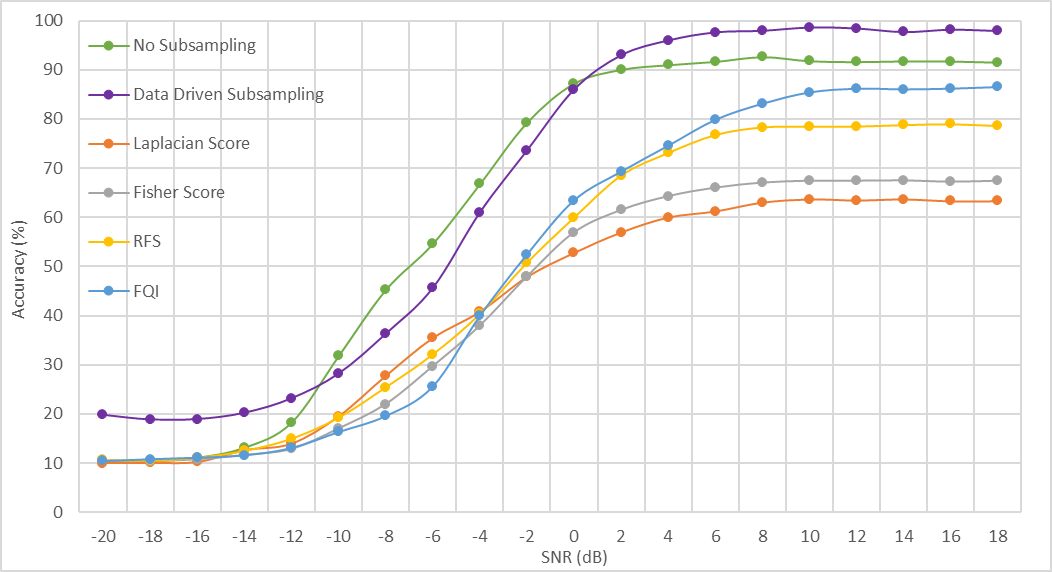}}}
    \caption{Accuracy vs SNR comparisons of the proposed Ensemble Wrapper Data-Driven Subsampler with (a) Conventional Subsampling Techniques and (b) Feature Selection Techniques for the ResNet classifier at $\frac{1}{8}$ subsampling rate.}
	\label{fig:resnet_comparisons}
\end{figure}

Feature selection algorithms aim at identifying important input vector elements. In the considered setup, a direct application of a feature selection algorithm for $2 \times 128$ input vectors, would treat each of the $256$ elements separately. We handle this with a slight modification to tie the real and imaginary parts of each sample, as illustrated below. In Figure \ref{fig:resnet_comparisons}b, we show a comparison between the proposed method and four popular feature selection algorithms; namely: 1) Laplacian Score \cite{he2006laplacian}: which is an unsupervised filter feature selection algorithm that selects features with the objective of preserving the data manifold structure through a graph representation \cite{Li2017FeatureSA}, 2) Fisher Score \cite{gu2012generalized}: which is a supervised filter feature selection algorithm that selects features such that the features of samples within the same modulation type are similar while the features of samples belonging to other modulation types are as distinct as possible \cite{Li2017FeatureSA}, 3) Efficient and Robust Feature Selection (RFS) \cite{nie2010efficient}: which is a computationally efficient embedded feature selection method that exploits the noise robustness - through rotational invariance - property of the joint $\ell_{2,1}$-norm loss function \cite{ding2006r, Li_2017}, by applying the $\ell_{2,1}$-norm minimization on both the loss function and its associated regularization function, and 4) Feature Quality Index (FQI) \cite{de1997feature}: which is a wrapper feature selection algorithm that utilizes the output sensitivity of the considered model to changes in the input to rank features. FQI can be considered as a simplified version of our Subsampler Net that relies on the Mean Squared Error (MSE) loss instead of the model's loss and uses only the initial sample ranking that we use to select the first sample. For each of the techniques examined, except FQI where we have sample scores, we add the two scores obtained for the two features belonging to a sample to obtain a sample score before proceeding to rank the samples. We note how the proposed method delivers significantly better performance than all considered methods, and that this holds for all other considered subsampling rates not shown in the figure. Particularly, our method delivers better performance than FQI, which demonstrates the need for re-ranking samples after each iteration in the Subsampler Net described in Section \ref{sec:subsampler_nets}. Also, we note that this re-ranking does not require model retraining, unlike most other wrapper feature selection algorithms, which makes our method computationally feasible in a wide range of settings.

\subsection{Ablation Study}\label{sec:ablation}
\begin{figure}[H]
    \captionsetup[subfigure]{labelformat=empty}
    \centering
    \subfloat[(a)]{{\includegraphics[width=0.45\columnwidth]{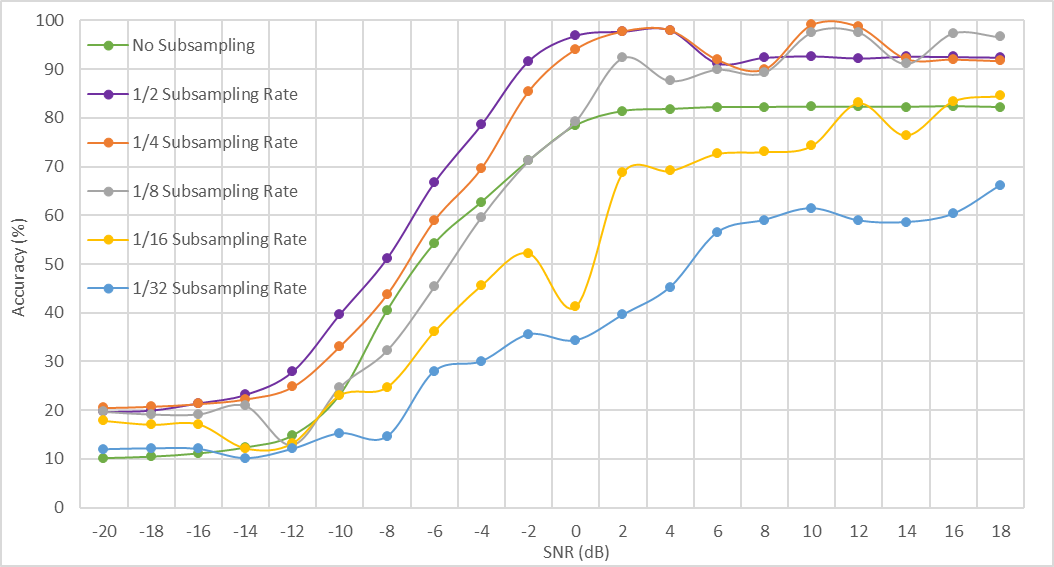}}}%
    \qquad
	\subfloat[(b)]{{\includegraphics[width=0.45\columnwidth]{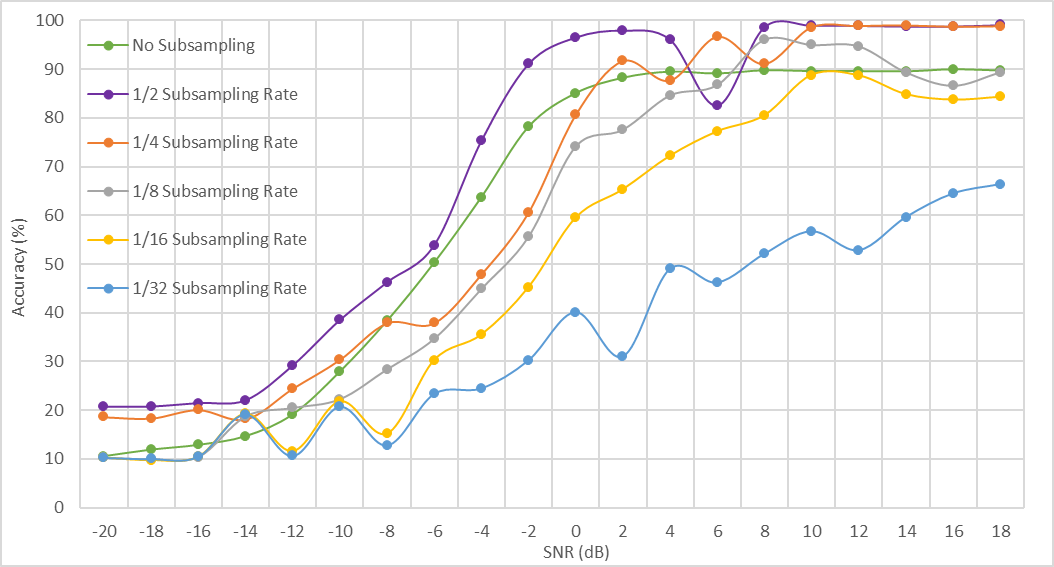}}}%
    \qquad
    \subfloat[(c)]{{\includegraphics[width=0.45\columnwidth]{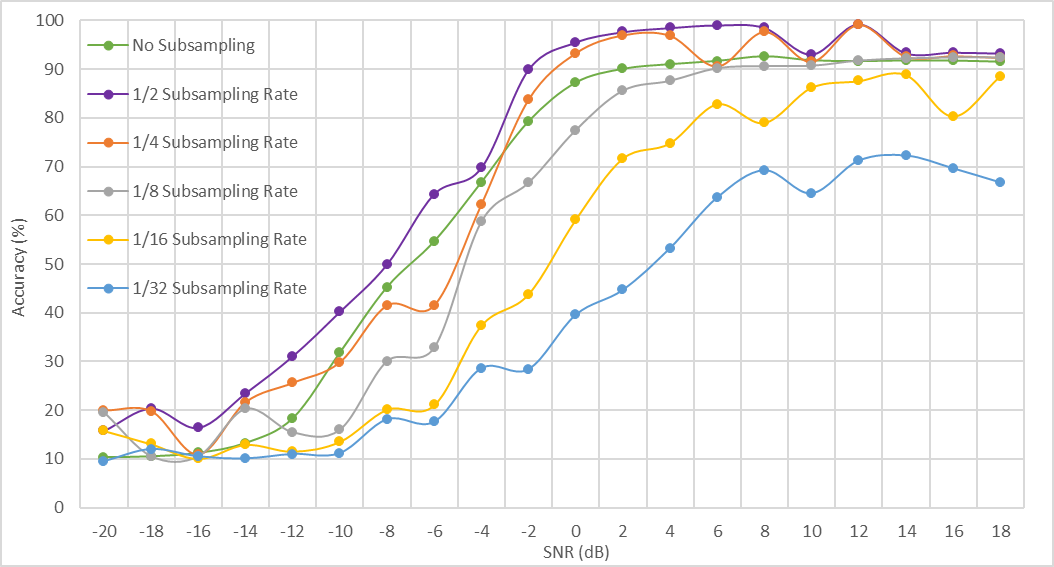}}}
    \qquad
    \subfloat[(d)]{{\includegraphics[width=0.45\columnwidth]{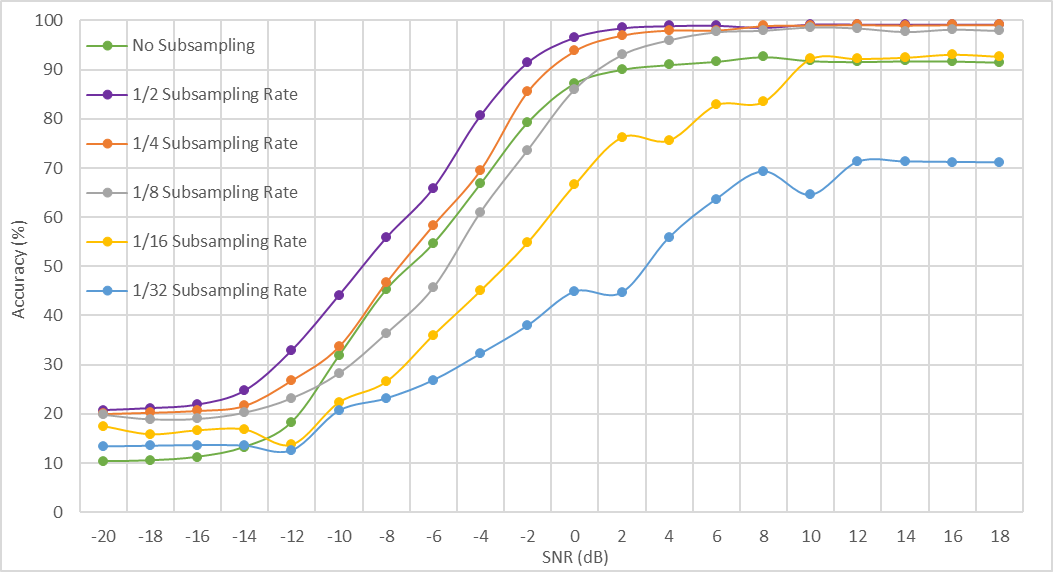}}}
    \caption{Accuracy vs SNR for ResNet classifier with (a) CNN Subsampler Net, (b) CLDNN Subsampler Net, (c) ResNet Subsampler Net, and (d) Holistic Subsampler.}
	\label{fig:ablation}
\end{figure}

We observe from Figure \ref{fig:ablation} how the relative performance of the three considered Subsampler Nets changes with different sampling rates and SNR values. This is the main motivation behind the Holistic Subsampler that benefits from the performance diversity among the three architectures. However, even the Holistic Subsampler, suffers from significant drops in classification accuracy for a wide range of SNR values at sampling rates well below the Nyquist rate. This motivated our $\epsilon$-Greedy step of the proposed approach. In particular, the slope of the classification accuracy curve for the Holistic Subsampler becomes negative towards -$12$, $4$, and $8$ dB with a $\frac{1}{16}$ subsampling rate and towards -$12$, $2$, and $10$ dB with a $\frac{1}{32}$ subsampling rate. To obtain the results shown in Figure \ref{fig:after_dds_plot}, our ensemble wrapper data-driven subsampling algorithm, illustrated in Algorithm \ref{algo:main}, applied $\epsilon$-Greedy Search with $\epsilon=\frac{1}{64}=\frac{2}{d}$ for the $4$ and $8$ dB SNR values with $\frac{1}{16}$ subsampling rate, as well as at $10$ dB SNR with $\frac{1}{32}$ subsampling rate. For the -$12$ dB SNR value with both $\frac{1}{16}$ and $\frac{1}{32}$ subsampling rates, an $\epsilon=\frac{1}{32}$ had to be used because having $\epsilon=\frac{1}{64}$ was insufficient, and the same held for $2$ dB SNR with $\frac{1}{32}$ rate. It is important to note that our $\epsilon$-Greedy step has a time complexity of the Order $O((\epsilon d)^k)$, if the number of explored combinations has the same order as the number of the constructed tree leaves. Fortunately, this step is typically needed only at very low subsampling rates, where the value of $k$ is small, and with small values of $\epsilon$.
\section{Discussion}\label{sec:discussion}
\subsection{Exploiting Transfer Learning}
We note how the Holistic Subsampler achieves better results than any individual Subsampler Net, even though the final classifier relies on a single ResNet architecture. Furthermore, even though each of these architectures is trained to classify the data when all the samples are present at the input, when used as Subsampler Nets, one or more of these samples are set to 0. Hence, we use the trained deep neural network classifiers in two ways other than their intended application that they are trained on: 1- They are used to select samples for another classifier, 2- They are used with only a subset of samples present. This is only possible because of the transferability property of these deep neural network architectures. In general, we believe that exploiting transferability has great potential for various wireless communication tasks that rely on processing received signal samples. 

\subsection{Subsampling leads to higher accuracy}\label{sec:subsampling_high_acc}
\begin{figure}
    \captionsetup[subfigure]{labelformat=empty}
    \centering
    \subfloat[PCA before subsampling]{{\includegraphics[width=0.30\columnwidth]{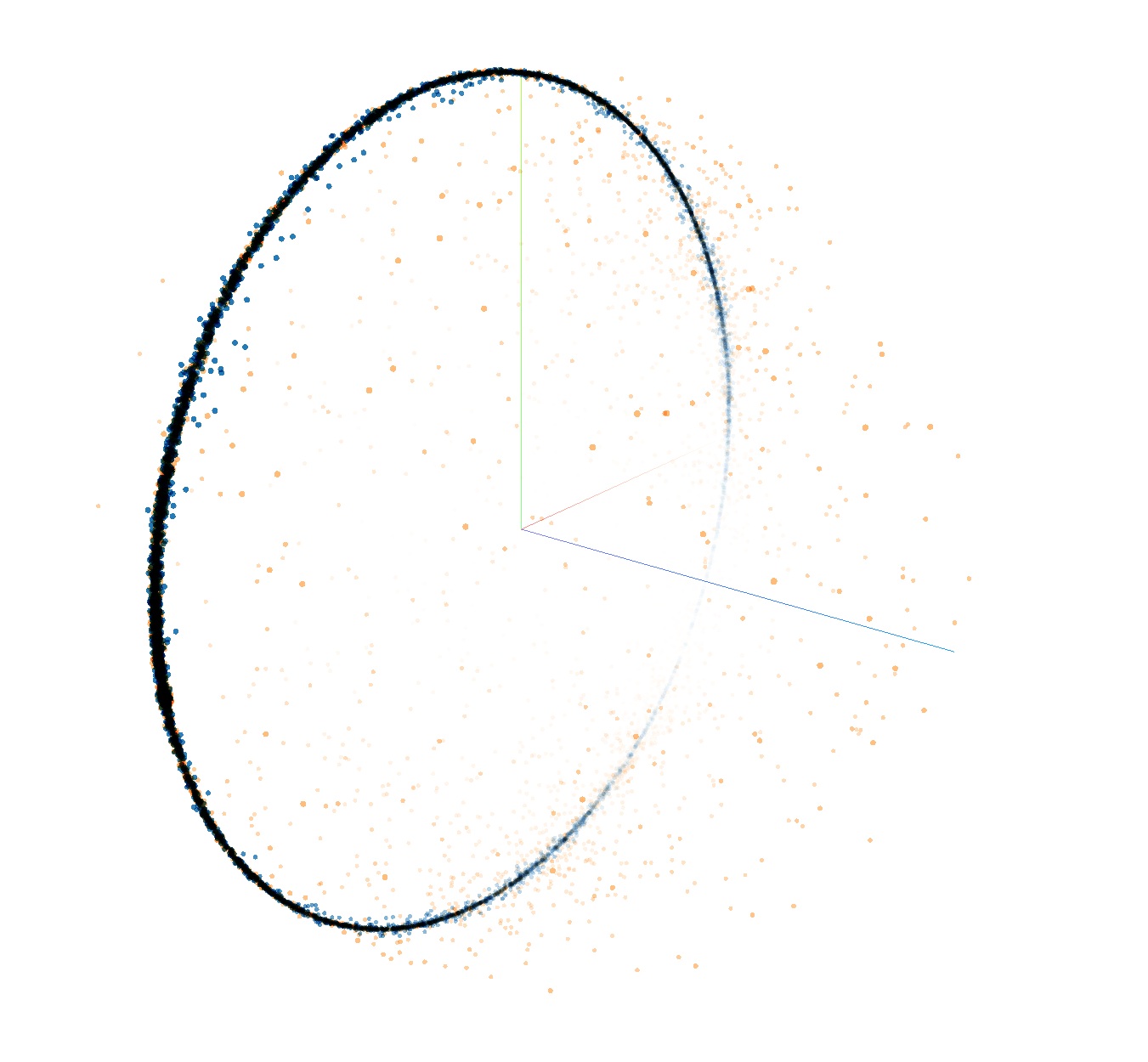}}}%
    \qquad \qquad \qquad 
    \subfloat[PCA after Ensemble Wrapper Subsampling]{{\includegraphics[width=0.30\columnwidth]{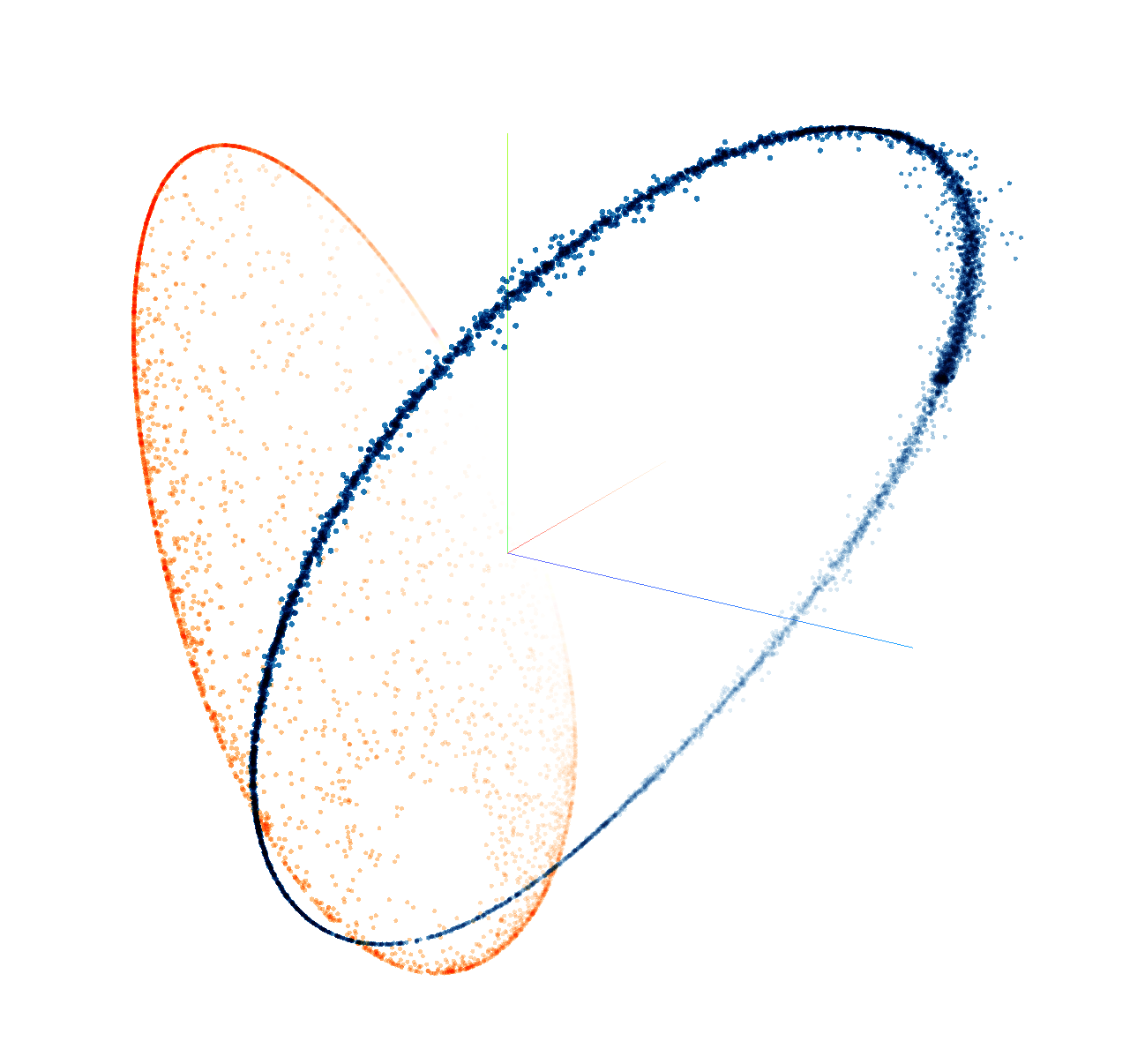}}}
	\qquad \qquad
	\subfloat[t-SNE before subsampling]{{\includegraphics[width=0.30\columnwidth]{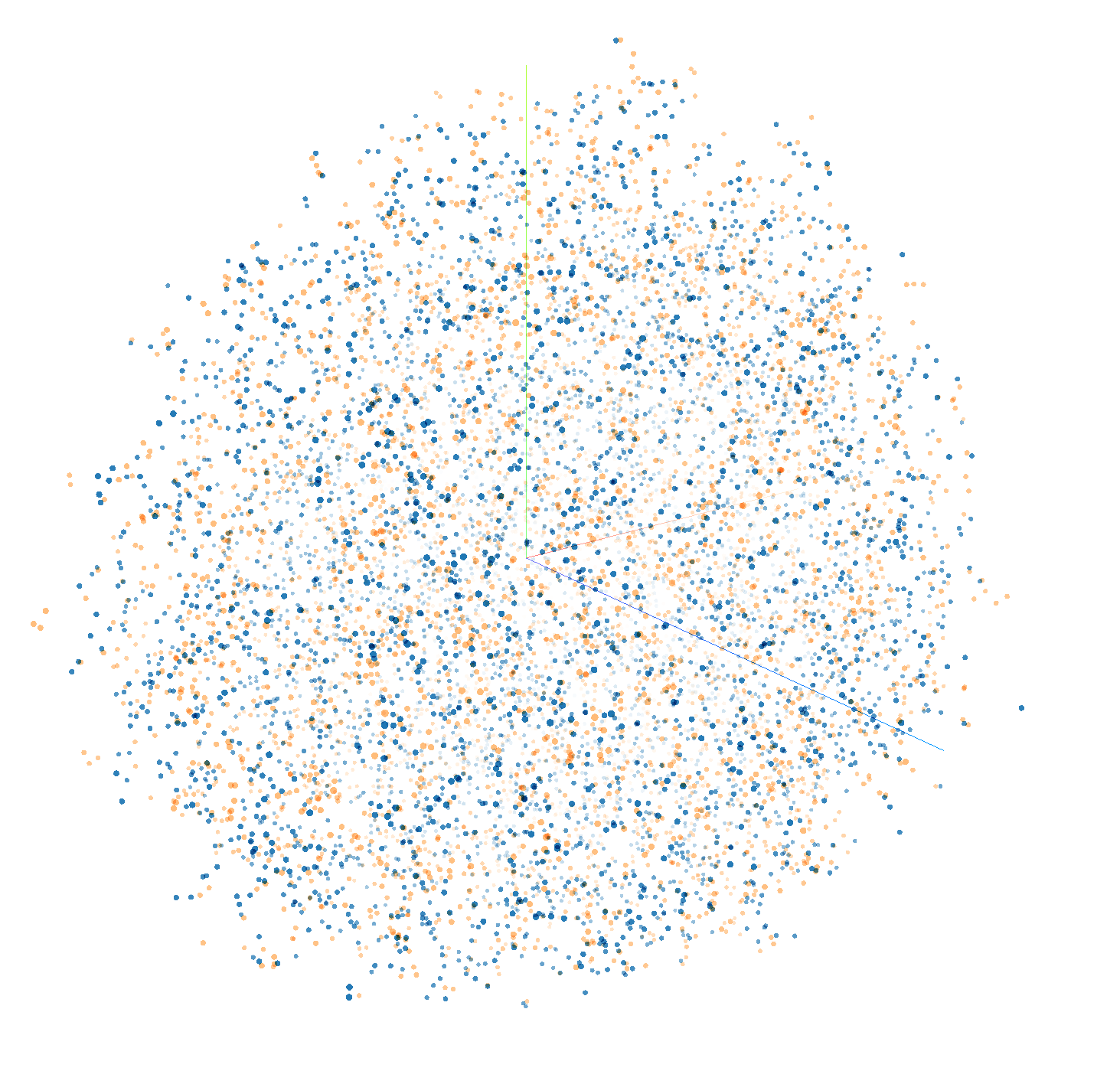}}}%
    \qquad \qquad \qquad 
    \subfloat[t-SNE after Ensemble Wrapper Subsampling]{{\includegraphics[width=0.30\columnwidth]{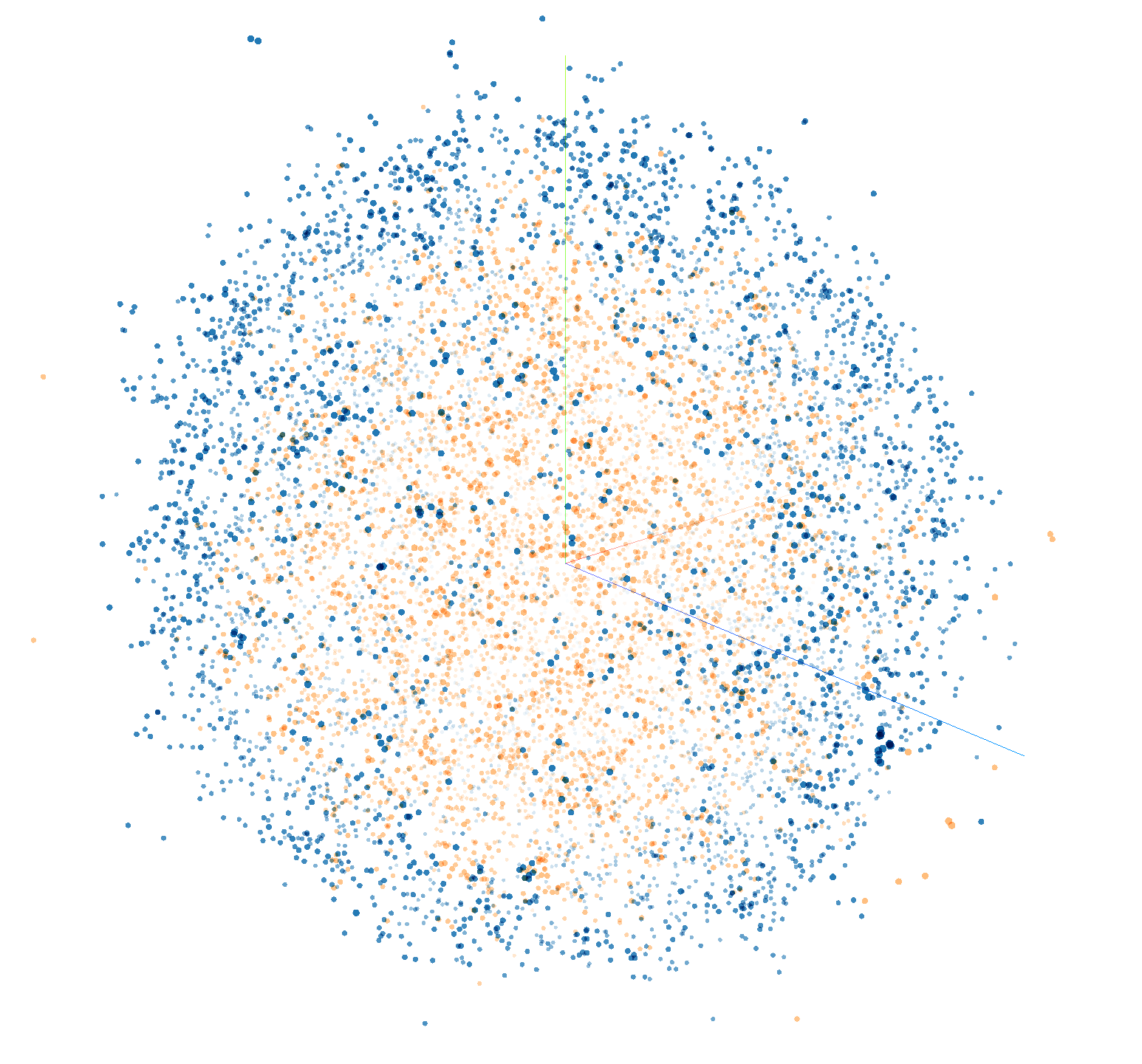}}}
    \caption{(Upper) PCA visualization of the training dataset for the AM-DSB (blue) and WBFM (orange) classes before and after Ensemble Wrapper Subsampling with a subsampling rate of $1/2$ at $18$ dB SNR. (Lower) t-SNE visualization for QAM16 (blue) and QAM64 (orange) at same rate and SNR.}
	\label{fig:pairs}
\end{figure}

In Section \ref{sec:nosubsampling}, we saw that all the deep learning architectures suffered from the same drawback of the AM-DSB/WBFM and QAM16/QAM64 misclassification when no subsampling is used. To further analyze why the proposed subsampling method leads to higher classification accuracies with fewer samples, we will be using Principal Component Anaysis (PCA) \cite{ivosev2008dimensionality} and t-Distributed Stochastic Neighbor Embedding (t-SNE) \cite{maaten2008visualizing} to visualize how subsampling allows us to reduce overfitting, particularly for the aforementioned class pairs. 
We first subsample the training dataset at $18$ dB SNR with a rate of $1/2$. After subsampling, we have $64$ samples, corresponding to $128$ features. Finally, we implement PCA and t-SNE to obtain a $3$-Dimensional projection of the training dataset for better visualization. We chose to implement both PCA and t-SNE because PCA clarifies the distinction between AM-DSB and WBFM while t-SNE clarifies the distinction between QAM16 and QAM64 after subsampling, as shown in Figure \ref{fig:pairs}\footnote{The t-SNE plots were generated with a perplexity value of $20$, a learning rate of $10$, and were run for $250$ iterations.}. Observing the figure, we believe that the higher accuracy values stem from the subsampling strategy enabling simpler decision boundaries to distinguish, with high fidelity, between the considered class pairs, which improves generalization performance and reduces overfitting. 

\subsection{Designing the Ranker Models}
\begin{figure}[H]
    \captionsetup[subfigure]{labelformat=empty}
    \centering
    \subfloat[(a)]{{\includegraphics[width=0.24\columnwidth]{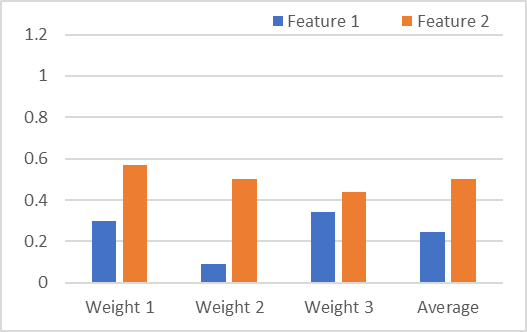}}}
    \qquad
    \subfloat[(b)]{{\includegraphics[width=0.24\columnwidth]{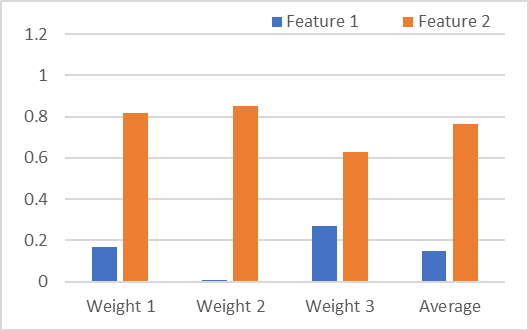}}}
    \qquad
    \subfloat[(c)]{{\includegraphics[width=0.24\columnwidth]{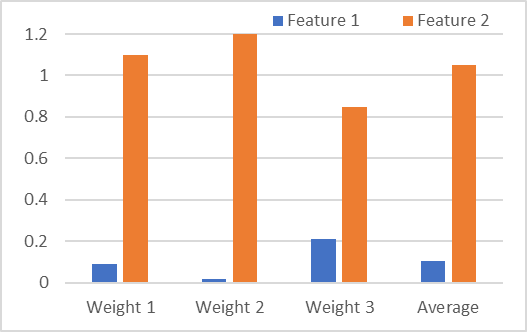}}}
    \caption{Input layer weight magnitudes for toy example after training for (a) 10, (b) 100, and (c) 1000 epochs.\vspace{-5mm}}
	\label{fig:toy_overfitting}
\end{figure}

The performance of a Subsampler Net heavily depends on the performance of the model used to rank the features. In some cases, however, even the state-of-the-art models do not have high classification accuracy values. In such cases, we believe that better feature selection results can be obtained with a Subsampler Net by training the ranker model for more epochs beyond what is suggested by the Early Stopping algorithm (see e.g., \cite[Chapter $8$]{dl-book}). This is as we found this strategy to be useful in multiple settings and further observed that it can significantly increase the discrepancy in weight magnitudes across the different features. For example, we considered a toy example constructed in TensorFlow Playground with a single-hidden-layer network that distinguishes between two classes based on two features, and the second is more salient as it enables forming a decision boundary that allows for better classification.
Each of these features have three weights in the input layer and as expected, the weights connected to Feature $2$ quickly manifest, after several epochs, into weights of higher average magnitude than those belonging to Feature $1$ as shown in Fig. \ref{fig:toy_overfitting}.
As the number of training epochs increases - even from $100$ to $1000$ which is way more than needed for this small network - the difference in the average magnitudes of the weights increases. 
This implies that the ranker will be able to better rank the saliency of features because the accuracy difference will increase when suppressing each of these features.

\subsection{Sensitivity to SNR estimate}
Even though both the final classifier and the ranker models used in the proposed subsampling method are trained using the whole training set across all considered SNR values, the selected set of sample indices is different for each SNR value, and hence, we expect a real time system employing this method to have an accurate estimate of the SNR value, in order to know the right set of sample indices. We made this choice, as we found it to deliver a significantly superior performance to the extreme alternative, where the same set of sample indices is selected for all SNR values. In future work, we plan to investigate the impact of small errors in such an estimate, by comparing the different sets of selected sample indices for adjacent SNR values. We plan to also benefit from analyzing these sets of sample indices, to better understand the roles of different ranker models at different SNR values.

\section{Concluding Remarks}\label{sec:conclusion}
In this work, we considered the problem of recognizing one out of ten modulation types with a constraint on the sampling rate in an erroneous wireless environment that is difficult to model. We first identified three deep neural network architectures that are well fit for the task and deliver state-of-the-art classification accuracy, namely a CNN, CLDNN and ResNet. We then presented a wrapper data-driven subsampling approach that employs all three architectures - as an ensemble - for selecting a set of samples that maximizes the classification accuracy via recursive simulations aided by $\epsilon$-Greedy deterministic explorations. Our experimental results, using the RadioML2016.10b dataset of \cite{conv}, indicate that using the proposed method with a ResNet classifier leads to very high classification accuracy values, that to the best of our knowledge, have not been reached before even at sampling rates well above the Nyquist rate. Further, even in the sub-Nyquist regime, we achieve almost perfect classification (accuracy above 99\%) at high SNR. We also noted the drastic reduction in the classifier's training time as a result of subsampling. We plan to further investigate in future work the potential of employing deep learning for subsampling in wireless communication systems, as we believe that the insights distilled from this work carry practical significance beyond the considered modulation classification task.

\bibliography{refs}
\bibliographystyle{IEEEtran}

\end{document}